# The effect of a band gap gradient on the radiative losses in the open circuit voltage of solar cells

**Sevan Gharabeiki** [1], Francesco Lodola [1], Tilly Schaaf [1,2], Taowen Wang [1], Michele Melchiorre [1], Nathalie Valle [3], Jérémy Niclout [3], Manha Ali [4], Yucheng Hu [4], Gunnar Kusch [4], Rachel A. Oliver [4] and Susanne Siebentritt [1]

1. Laboratory for Photovoltaics, Department of Physics and Material Science, University of Luxembourg, 41, rue du Brill, L-4422, Belvaux, Luxembourg

2. On leave from Trinity College Dublin, College Green, Dublin 2 Ireland

3. Luxembourg Institute of Science and Technology (LIST), 41, rue du Brill, L-4422, Belvaux, Luxembourg

4. Department of Materials Science and Metallurgy, University of Cambridge, 27 Charles Babbage Road, Cambridge CB3 0FS, United Kingdom

Email: sevan.gharabeiki@uni.lu, susanne.siebentritt@uni.lu

**Keywords**: Photoluminescence, Electroluminescence, Quasi-Fermi Level Splitting, Generalized Detailed Balance Model, Voltage Loss Analysis, Chalcopyrite, Meta-Analysis

## Abstract:

The radiative open-circuit voltage loss in a solar cell occurs because the absorptance spectrum near the band gap shows gradual increase rather than sharp step-function like transition. This broadening effect has been attributed to band gap fluctuations and/or to Urbach tails. In this report, we use modelling based on Planck's generalized law to distinguish between these two effects. Our results demonstrate that Urbach tails have only a minimal effect on the absorptance edge broadening and clarify that even an ideal direct semiconductor with no band gap fluctuations shows broadening at the absorptance onset. Furthermore, state-of-the-art inorganic thin-film solar cells often incorporate a band gap gradient across their thickness, which can further contribute to absorptance broadening. Using Cu(In,Ga)Se$_2$ (CIGSe) absorbers as a case study, we perform a comprehensive analysis of voltage losses through absolute photoluminescence and electroluminescence spectroscopy, combined with photospectrometry and high-spatial-resolution cathodoluminescence measurements. We find that the loss analysis based on the combination of radiative, generation and non-radiative losses is complete. Samples with a graded band gap profile show more pronounced broadening of the absorptance onset and up to 16 mV higher radiative losses compared to the samples with uniform band gap. There is indication, that band gap-graded samples also have larger lateral band gap inhomogeneity.

# 1. Introduction

The quasi-Fermi level splitting ($\Delta\mu$) in a photovoltaic device is analogous to the open circuit voltage ($V_{OC}$) [1-5]. The Shockley-Queisser model predicts the maximum achievable $\Delta\mu$ (denoted here as $qV_{OC}^{SQ}$) of an idealized solar cell with a step-like absorptance *A(E)* spectrum [6], where *A(E)* is zero below the band gap and one above it. On the other hand, real solar cells exhibit a more or less gradual increase in the *A(E)* edge [7]. Multiple factors affect the broadening of the *A(E)* spectra, for instance, presence of lateral bandgap fluctuations [8-10], sub band gap states [11], thickness of the absorber layer [4, 12] and presence of compositional gradient in the absorber [9, 12] can all contribute to the broadening of the *A(E)* profile. Additionally, this broadening is more pronounced in indirect semiconductors than in direct ones [7].

For a solar cell with a gradual *A(E)* onset, the maximum attainable $\Delta\mu$, known as the radiative $\Delta\mu$ limit ($\Delta\mu^{rad}$), has lower values than the SQ limit ($qV_{OC}^{SQ}$) [4, 7, 8, 13]. The difference is due to two effects: (i) increased emission, because lower energy states contribute more strongly to the radiative emission [4] (radiative loss $\delta\Delta\mu^{rad}$) and (ii) lower generation, mostly because of reflection losses (generation loss $\delta\Delta\mu^{Gen}$). Further losses are due to non-radiative recombination (i.e. Shockley-Read-Hall [14, 15] and Auger-Meitner recombination [16-18]).

This separation of the different loss mechanisms in $V_{OC}$ was first introduced by Rau et al. [7]. Their approach was based on the optoelectronic reciprocity principle [19], which links the electroluminescence (EL) spectrum of any solar cell to its short circuit current quantum efficiency (*QE(E)*) spectrum. They introduced the radiative $V_{OC}$ loss ($\delta V_{OC}^{rad}$), based on the increased radiative emission due to the gradual increase of the *QE(E)* spectrum, the short-circuit $V_{OC}$ loss ($\delta V_{OC}^{SC}$), due to a decrease in the short-circuit current density (*J$_{SC}$*) of the solar cell caused by incomplete absorption and charge collection losses, and the non-radiative $V_{OC}$ loss ($\delta V_{OC}^{nr}$), due to non-radiative recombination. In their approach the non-radiative losses were determined from the difference of the measured $V_{OC}$ and its radiative $V_{OC}$ limit ($V_{OC}^{rad}$). However, absolute PL measurements allows to determine non-radiative losses directly from the PL quantum efficiency (*Y$_{PL}$*). Therefore, here we analyze the different losses individually from PL and photospectrometric absorptance measurements and compare them to the $\Delta\mu$ determined from a linear fit to generalized Planck's law (analogous to *V$_{OC}$*) [3, 4, 20, 21] to investigate if the loss analysis is complete and adds up. We start by describing the analogies and the differences between the two approaches, based on the optoelectronic reciprocity principle and based on generalized Planck's law. Then we apply generalized Planck's law to a couple of model systems that allow us to separate the influence of the band gap fluctuations and Urbach tails, as well as link real band gap distributions to the distribution of step-function band gaps.

State-of-the-art inorganic thin film solar cells usually incorporate a band gap gradient profile across the absorber thickness to minimize interface recombination. In Cu(In,Ga)Se$_2$ (CIGSe) solar cells, this gradient is achieved by increasing Ga concentration toward both the front and back surfaces, resulting in a higher band gap at each end [11, 22-25]. The presence of higher Ga concentration (i.e., higher band gap) near the back contact is essential to mitigate back surface recombination [26, 27]. Additionally, the front gradient helps to reduce front interface recombination if this interface is not ideal [26]. In Cd(Se,Te) solar cells, this band gap engineering involves changing the Se and Te composition through the depth [28, 29]. In emerging technologies, such as CuZnSn(S,Se)$_4$ there are also attempts to engineer the band gap along the absorber thickness. Particularly, adjustments in S and Se concentration across the depth yielded efficiencies up to 13.7% by using a "V-shaped" band gap profile [30]. The presence of a gradient softens the absorption edge further, since the region with the minimum bandgap extends only over a short distance in the absorber depth, not providing complete absorption of the near edge light. To clarify the influence of a band gap gradient on the radiative voltage loss, we experimentally investigate several CIGSe films and solar cells with and without band gap gradient, using both approaches: based on reciprocity and based on generalized Planck's law. We will show in the following (i) that the loss analysis introduced by Rau et al. [7] is complete, i.e. the losses add up to the actual $\Delta\mu$, and (ii) that the band gap gradient softens the absorption edge, i.e. a gradient increases the width of the bandgap distribution by about 10 meV to 20 meV and the radiative loss by 5 mV to 16 mV.

Furthermore, by performing a meta-analysis of published CIGSe data, we demonstrate that the broadening of $A(E)$ is directly correlated with the strength of the Ga gradient throughout the depth. Our analysis reveals that when the minimum band gap region is confined to a narrow thickness, the absorptance edge is broader compared to the cases where the minimum band gap extends significantly through the depth of the absorber.

## 2. Theory and generalized detailed balance model

The radiative $V_{OC}$ loss ($\delta V_{OC}^{rad}$) was first introduced by Rau et al [7], In their study, they expanded the loss analysis to also include the short circuit $V_{OC}$ loss ($\delta V_{OC}^{SC}$) and the non-radiative $V_{OC}$ losses ($\delta V_{OC}^{nr}$). We refer to this analysis as the "generalized detailed balance model". In its original version the model is derived from the optoelectronic reciprocity principle [7, 19], and relies on the optoelectronic characterization of complete solar cells by the external quantum efficiency *QE(E)* spectrum. This model can be expanded to purely optical photoluminescence measurements based on solar absorbers, without finishing the solar cell structure [4]. This purely optical approach makes use of generalized Planck's law as the starting point. In **Table 1** we directly compare the formalism based on the reciprocity principle (optoelectronic measurements) to the one based on generalized Planck's law (optical measurements). For simplicity, all symbols and constants used in this study are listed in the **Appendix A** at the end of this manuscript. Detailed derivation of the loss analysis based on optical characterization (i.e., using Planck's generalized law) is presented in **Appendix B**.

Generalized Planck's law (***Eq*. 1**) is based on optical measurements. It links the PL emission flux spectrum $\Phi_{PL}(E)$ to the absorptance spectrum *A(E)* and allows to analyze the losses of the $\Delta\mu$ of an absorber before finishing the solar cell. This approach is completely analogous to the optoelectronic analysis based on the reciprocity principle, which links the electroluminescence flux spectrum $\Phi_{EL}(E)$ to the external quantum efficiency spectrum *QE(E)* and allows to analyze the losses of the open circuit voltage $V_{OC}$ [19]. The optical approach allows us to generalize the detailed balance model for non-step like *A(E)* spectra and determine the $\Delta\mu^{rad}$ for an absorber material with any given *A(E)* profile even without finishing the solar cell [4].

Table 1 Summarized key equations for generalized detailed balance model, based on both optical and optoelectronic measurements. (Symbols and constants are explained in **Appendix A** at the end of this manuscript)

| | **Optical measurements** | **Optoelectronic measurements** |
|---|---|---|
| **Equation to start with** | Generalized Planck's law $$\Phi_{PL}(E) \approx A(E)\Phi_{BB}(E)\exp\left(\frac{\Delta\mu}{kT}\right) \ (Eq.1)$$ | Reciprocity principle $$\Phi_{EL}(E) \approx QE(E)\Phi_{BB}(E)\exp\left(\frac{V_{in}}{kT}\right) \ (Eq.7)$$ |
| **Radiative limit** | $$\Delta\mu^{rad} = k_BT\ln\frac{\int_0^\infty A(E)\Phi_{Sun}dE}{\int_0^\infty A(E)\Phi_{BB}(E)dE}$$ $$= k_BT\ln\frac{F_{Gen}^{Sun}}{F_0^{rad}} \ (Eq.2)$$ | $$V_{OC}^{rad} = \frac{k_BT}{q}\ln\frac{\int_0^\infty QE(E)\Phi_{Sun}dE}{\int_0^\infty QE(E)\Phi_{BB}(E)dE}$$ $$= \frac{k_BT}{q}\ln\frac{J_{SC}}{J_0^{rad}} \ (Eq.8)$$ |
| **Radiative loss** | $$\delta\Delta\mu^{rad} = -k_BT\ln\left(\frac{\int_{E_g}^\infty \Phi_{BB}(E)dE}{\int_0^\infty A(E)\Phi_{BB}(E)dE}\right)$$ $$= -k_BT\ln\left(\frac{F_0^{SQ}}{F_0^{rad}}\right) \ (Eq.3)$$ | $$\delta V_{OC}^{rad} = -\frac{k_BT}{q}\ln\left(\frac{\int_{E_g}^\infty \Phi_{BB}(E)dE}{\int_0^\infty QE(E)\Phi_{BB}(E)dE}\right)$$ $$= -\frac{k_BT}{q}\ln\left(\frac{J_0^{SQ}}{J_0^{rad}}\right) \ (Eq.9)$$ |
| **Generation/short-circuit loss** | $$\delta\Delta\mu^{Gen} = -k_BT\ln\left(\frac{\int_0^\infty A(E)\Phi_{Sun}dE}{\int_{E_g}^\infty \Phi_{Sun}dE}\right)$$ $$= -k_BT\ln\left(\frac{F_{Gen}^{Sun}}{F_{Gen}^{SQ}}\right) \ (Eq.4)$$ | $$\delta V_{OC}^{SC} = -\frac{k_BT}{q}\ln\left(\frac{\int_0^\infty QE(E)\Phi_{Sun}dE}{\int_{E_g}^\infty \Phi_{Sun}dE}\right)$$ $$= -\frac{k_BT}{q}\ln\left(\frac{J_{SC}}{J_{SC}^{SQ}}\right) \ (Eq.10)$$ |
| **Non-radiative losses** | $$\delta\Delta\mu^{nr} = -k_BT\ln(Y_{PL}) \ (Eq.5)$$ | $$\delta V_{OC}^{nr} = -\frac{k_BT}{q}\ln(Y_{EL}) \ (Eq.11)$$ |
| **Final $\Delta\mu$ /$V_{in}$** | $$\Delta\mu = \Delta\mu^{rad} - \delta\Delta\mu^{nr}$$ $$= qV_{OC}^{SQ} - \delta\Delta\mu^{rad} - \delta\Delta\mu^{Gen} - \delta\Delta\mu^{nr} \ (Eq.6)$$ | $$V_{in} = V_{OC}^{rad} - \delta V_{OC}^{nr}$$ $$= V_{OC}^{SQ} - \delta V_{OC}^{rad} - \delta V_{OC}^{SC} - \delta V_{OC}^{nr} \ (Eq.12)$$ |

***Eq. 6*** describes how $\Delta\mu^{rad}$ deviates from the ideal SQ limit ($qV_{OC}^{SQ}$). Here, $\delta\Delta\mu^{rad}$ represents the radiative $\Delta\mu$ loss (***Eq. 3***), while $\delta\Delta\mu^{Gen}$ represents the generation $\Delta\mu$ loss (***Eq. 4***).

The $\delta\Delta\mu^{rad}$ is due to the broadening of the absorption edge (i.e., non-step like *A(E)*). It depends on the radiative saturation flux density, $F_0^{rad}$ (See ***Eq. 3***). When *A(E)* broadens, the integral over *A(E)* multiplied by $\Phi_{BB}$ increases due to the strong increase of the black body spectrum $\Phi_{BB}$ towards smaller energies, leading to higher values of $F_0^{rad}$ compared to the SQ radiative saturation flux densities ($F_0^{SQ}$), which is based on a step-function absorption. This trend increases the $\delta\Delta\mu^{rad}$. Descriptively speaking: the radiative $\Delta\mu^{rad}$ is the quasi-Fermi level splitting that occurs at the balance between generation flux and radiative emission flux. Because the emission flux increases when lower energy states are involved, a lower quasi-Fermi level splitting is needed to balance the incoming generation flux (see ref [4] and **Appendix B**).

Furthermore, the generation loss ($\delta\Delta\mu^{Gen}$) occurs due to incomplete absorption, since the *A(E)* does not reach unity for energies above the band gap due to factors like reflection losses or insufficient thickness. Therefore, incomplete absorption reduces the sun generation flux ($F_{Gen}^{Sun}$) below the SQ generation flux ($F_{Gen}^{SQ}$) (See ***Eq. 4***). This loss mechanism simply means that fewer photons are absorbed, resulting in a smaller number of photogenerated charge carrier generation contributing to $\Delta\mu$.

Additionally, the most important loss mechanism in real absorbers is the non-radiative $\Delta\mu$ loss ($\delta\Delta\mu^{nr}$) (***Eq. 5***), which reduces the final $\Delta\mu$ below the radiative limit ($\Delta\mu^{rad}$) (***Eq. 6***). This loss mechanism can be determined using photoluminescence quantum yield ($Y_{PL}$) of the absorber. The $Y_{PL}$ is defined as the ratio of emitted photon flux out of the absorber over the absorbed generation flux [4, 31]:

$$Y_{PL} = \frac{\int_0^\infty \Phi_{PL}(E)\,dE}{F_{Gen}^{Sun}} \approx \frac{\int_0^\infty \Phi_{PL}(E)\,dE}{F_{Laser}^{Gen}} \qquad (Eq.\,13)$$

The second part of the equation describes the experimental measurement condition in our laboratory, where for PL measurements we use a monochromatic laser light source. Here, $F_{Laser}^{Gen} = A(E_{laser})F_{Laser}^{inc}$, with $F_{Laser}^{inc}$ is the incoming laser flux. More details regarding the calibration of PL measurements and laser fluxes are explained in the **experimental section** and **Supplementary information (SI) note 1.**

In an ideal absorber the $Y_{PL}$ is 1, which translates to no non-radiative recombination activities. However, in the real case $Y_{PL}$ is much smaller than 1 [3].

In analogy with generalized Planck's law of radiation (**Eq. 1**), the optoelectronic reciprocity relation (**Eq. 7**) makes a connection between EL emission signal $\Phi_{EL}(E)$ and the *QE(E)* spectrum and internal voltage ($V_{in}$) of a solar cell [19]. Here, $V_{in}$ is completely analogous to the $\Delta\mu$. In general it is different from the applied bias voltage because of transport losses at contacts and in the p/n junction [32]. Notably, the *A(E)* and *QE(E)* spectra are interconnected through the charge collection function [33] (See **SI note 2**). In a solar cell with an efficient charge collection, where all photogenerated charge carriers are extracted through the junctions, the *A(E)* would be equal to *QE(E)*.

Similar to $\Delta\mu$ losses, we define the voltage losses in **Eq. 12**, following reference [7]. Here, the $\delta V_{OC}^{rad}$ arises due to *QE(E)* onset broadening and $\delta V_{OC}^{SC}$ is due to incomplete absorption and incomplete charge collection. Similar to **Eq. 13**, we can define the electroluminescence quantum yield $Y_{EL}$, which is given by the ratio of the emitted flux density (i.e. $\int_0^\infty \Phi_{EL}(E)\,dE$) over the injected current density:

$$Y_{EL} = \frac{q \int_0^\infty \Phi_{EL}(E)\, dE}{J_{SC}} \qquad (Eq.\,14)$$

For absolute EL measurements under one sun condition, we need to inject current densities equal to short circuit current density ($J_{SC}$) [34].

## 3. Effect of absorption edge broadening and Urbach tails on $\delta\Delta\mu^{rad}$

The *A(E)* edge broadening in an absorber can be attributed to several factors. In a polycrystalline compound semiconductor such as CIGSe, alloy disorder arising from the presence of indium and gallium, as well as fluctuations in copper content, can introduce inhomogeneities in the band gap energy [9, 10]. These band gap fluctuations can occur both between different grains and within every individual grain, they occur intentionally or unintentionally, in any case leading to a broadening of the *A(E)* onset [9, 10, 35-37]. Additionally, electrostatic potential fluctuations, dislocations, grain boundaries, strain, and sub-band gap absorption (i.e., Urbach tails) further influence *A(E)* edge broadening [9, 10, 12, 35]. Overall, in polycrystalline material such as CIGSe these factors are often interconnected and cannot be easily separated, collectively contributing to the *A(E)* edge broadening. Furthermore, high-efficiency CIGSe absorbers are prepared with an intentionally graded band gap profile within their thickness [22, 23, 38, 39]. This gradient involves an increase in the Ga content towards the front and back surfaces. In this case the effective absorption thickness for energies near the absorption edge is confined to the narrow range of the minimum band gap position. This small spatial extension contributes to the broadening of the absorption edge. Moreover, the absorption coefficient *α(E)* at any given energy above the absorption edge in CIGSe absorbers decreases with increasing Ga concentration [33, 40, 41]. Consequently, with the presence of a gradient, *α(E)* at any given energy above the bandgap decreases towards the front and back surfaces, thereby contributing additionally to the broadening of the *A(E)* edge [9]. In this section, by considering a Gaussian distribution of band gaps, we study the effect of lateral band gap fluctuations and the presence of Urbach tails on the *A(E)* edge broadening and the radiative loss $\delta\Delta\mu^{rad}$. Importantly, we will demonstrate that the fluctuations and Urbach tails are distinct. However, both contribute to the broadening at the *A(E)* edge.

As proposed in earlier studies, the lateral band gap fluctuations across the absorber can be defined as Gaussian distribution of local band gaps ($E_g^{loc}$) around a mean value of the band gap energy ($\overline{E_g}$) [8, 9]:

$$P(E_g^{loc}) = \frac{1}{\sigma_g\sqrt{2\pi}}\exp\left(-\frac{(E_g^{loc}-\overline{E_g})^2}{2\sigma_g^2}\right) \qquad (Eq.\,15)$$

Here $\sigma_g$ is the band gap broadening parameter which is a measure to the degree of the band gap inhomogeneities within the absorber.

Considering lateral fluctuation, at each point on the absorber surface, we can define the local absorptance spectrum $A^{loc}(E, E_g^{loc})$. Subsequently, the overall *A(E)* spectrum can be obtained by integrating the $A^{loc}(E, E_g^{loc})$ over the probability distribution of band gaps [8, 9, 34], thus we can write:

$$A(E) = \int_0^\infty A^{loc}(E, E_g^{loc}) P(E_g^{loc}) dE_g^{loc} \qquad (Eq.\,16)$$

For the first approximation, we assume that at each point on the absorber surface the local absorptance is a step function (i.e., $A^{loc}(E, E_g^{loc}) = 1$ for $E > E_g^{loc}$ and $A^{loc}(E, E_g^{loc}) = 0$ for $E < E_g^{loc}$) [8, 9, 34]. In this case, the integration over a Gaussian distribution function would result in a complementary error function for the *A(E)* spectrum [8, 9, 34]:

$$A(E) = \frac{1}{2}\text{erfc}\left(\frac{\overline{E_g} - E}{\sqrt{2}\sigma_g}\right) \qquad (Eq.\,17)$$

For any *A(E)* spectrum, the first derivative of the *A(E)* spectrum can be used to evaluate the *A(E)* edge broadening with respect to step-like band gaps [7]. Often this derivative can be fitted by a Gaussian function with a broadening $\sigma_A$. It is obvious that for this simplified model (i.e, error function model in **Eq. 17**), the first derivative of *A(E)* would result in a Gaussian distribution with broadening parameter same as $\sigma_g$ ($\sigma_A = \sigma_g$), and with maximum peak position at the average band gap $\overline{E_g}$ [8, 9]. $\overline{E_g}$ is the bandgap we use to determine $V_{OC}^{SQ}$ (see **Fig C1** in **Appendix C**). In **Appendix C** we also performed some model calculations based on this simplified model where we calculated the $\Delta\mu^{rad}$ and $\delta\Delta\mu^{rad}$ using **Eq. 2** and **Eq. 3.** We show that with increase in $\sigma_g$, the $\delta\Delta\mu^{rad}$ also increases significantly. Furthermore, we compared our numerical calculations with those reported in previous publications [8] and found that our results are consistent with them (**Fig. C2** in **Appendix C**).

In the next section, we will show that this model provides a very useful approximation for experimentally measured *A(E)* spectra. Even though this approach is strictly valid only for lateral bandgap fluctuations, it provides also a useful analysis tool for bandgap variations in the depth of the absorber, like in graded samples.

Although assuming the error function like model for *A(E)* provides a useful and simple approximation for *A(E)*, this model does not offer any insights into the impact of Urbach tails. To address this issue, we employ a more accurate approximation for $A^{loc}(E, E_g^{loc})$, based on the absorption coefficient of a direct semiconductor above the bandgap and an Urbach tail below the bandgap. We assume at each point on the absorber, for a well-defined local band gap $E_g^{loc}$, the absorption coefficient ($\alpha$) is [34, 42]:

$$\alpha(E, E_g^{loc}) = \begin{cases} \alpha_0 \sqrt{\dfrac{E - E_g^{loc}}{k_B T}} & , \quad E > E_g^{loc} + \dfrac{E_U}{2} \\ \alpha_0 \exp\left(\dfrac{E - E_g^{loc}}{E_U}\right) \sqrt{\dfrac{E_U}{2e k_B T}} & , \quad E < E_g^{loc} + \dfrac{E_U}{2} \end{cases} \quad (Eq.\,18)$$

Where the $E_U$ is Urbach energy and $\alpha_0$ is a constant. Here, we use $\alpha_0 = 10^4$ cm$^{-1}$, which is consistent with values reported for most inorganic direct semiconductors [33, 43]. In this equation, we use a square root behavior (SQRT) for $\alpha$ at the energies above the $E_g^{loc}$ which is typical behavior of a direct semiconductor [44], and we use Urbach decay behavior for energies below it. The square root prefactor used in the Urbach decay term ensures continuity of the equation at the transition energy point. According to Beer-Lamberts law, at each point on the absorber surface we can define $A^{Loc}(E, E_g^{loc}) = 1 - \exp(-\alpha d)$, where the $d$ is the absorber thickness [34], then as before we integrate $A^{Loc}(E, E_g^{loc})$ over a Gaussian distribution of the band gap to extract the overall $A(E)$ (**Eq. 16**).

Similar approach was developed earlier by Redinger et al. to simulate the impact of thickness and Urbach tails on the $A(E)$ shape of a direct semiconductor [45]. Here, along with these two parameters, effect of lateral Gaussian fluctuations are also included in the simulations.

This model assumes that the sample exhibits two-dimensional lateral band gap fluctuations [34, 42], with each point on sample conceptualized as a vertical rod that absorbs light along its length. In **Fig 1. a**, the $A(E)$ spectrum for $\sigma_g = 40$ meV and $d = 2$ μm is illustrated for different $E_U$ values. In all our simulations we used the $\overline{E_g} = 1.1\ eV$ and a temperature of 296 K. $\overline{E_g}$ is the bandgap we use to determine $V_{OC}^{SQ}$. Similar to the error function model, the Gaussian fit to the derivative of $A(E)$ can be used as a measure of the $A(E)$ broadening ($\sigma_A$) (**Fig 1.b**). However, it should be noted that in this case the $\sigma_A$ is no longer the same as $\sigma_g$. Here, $\sigma_A$ has a more complex function, depending not only on the $\sigma_g$ but also on $E_U$ and the thickness d. **Fig 1.b** demonstrates a difference between $\sigma_g$ and $\sigma_A$. Even without band tails ($E_U = 0$ meV) $\sigma_A$ is larger than $\sigma_g$. To discuss this difference, we remind ourselves that the derivative of the $A(E)$ spectrum gives the distribution of SQ step-like band gaps (i.e., error function approach) and thus $\sigma_A$ approximates the width of the distribution of SQ type band gaps.

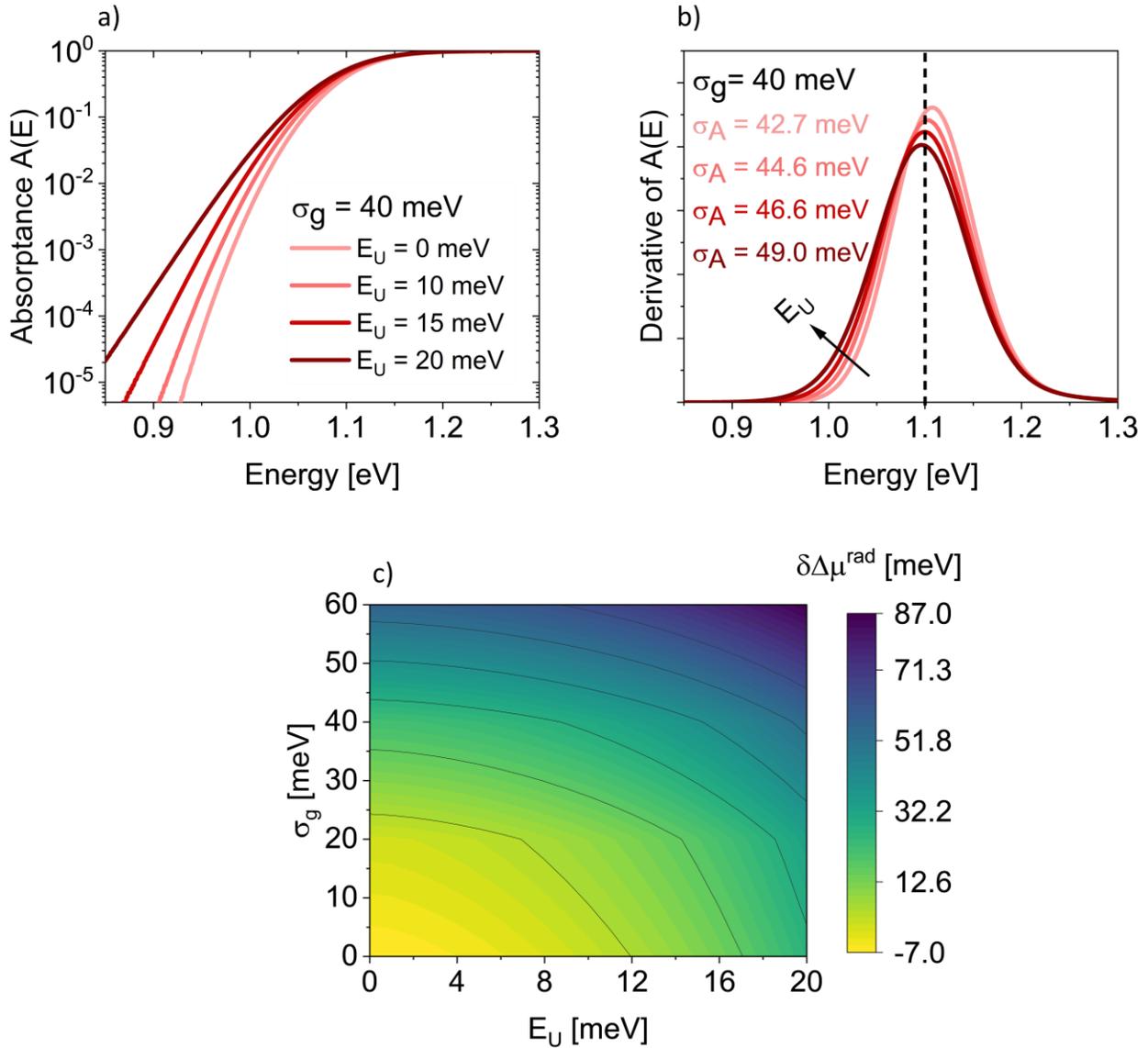

**Fig 1. a)** and **b)** represents the *A(E)* spectra and first derivative of *A(E)* for constant $\sigma_g$ = 40 meV and different Urbach energies, respectively. For each $E_U$, a distinct color code is assigned. In **b)** the $\sigma_A$ was extracted from a Gaussian fit to the first derivative of *A(E)*, and values are summarized for each derivative, the direction of the arrow shows increase in Urbach energy. **c)** Color map indicates the increase in the $\delta\Delta\mu^{rad}$ with increase in both $E_U$ and $\sigma_g$ values. (The $\delta\Delta\mu^{rad}$ was calculated using **Eq. 3** with $E_g = \overline{E_g} = 1.1 \, eV$ and T=296 K)

It is important to note that, even in the ideal case of a direct semiconductor, where the absorption coefficient shows square root behavior above the band gap (SQRT model), a distribution of SQ-type band gaps is needed. We demonstrate this in **Appendix D**, **Fig D1**. Since for a pure SQRT-type absorptance the derivative diverges at $\overline{E_g}$, we add a tiny tail of $E_U$= 0.01meV. The derivative is not Gaussian shaped, but when fit with a Gaussian, results in $\sigma_A$ of ~ 2 meV (See **Fig S2** in **SI**), although there is no distribution of ideal band gaps ($\sigma_g$=0). In the **SI note 3** we also have discussed the effect of the thickness on the *A(E)* edge broadening. As the absorber thickness decreases, the *A(E)* edge gets broader.

Additionally, as $E_U$ increases, the absorption of sub-band gap photons increases, leading to further broadening of the *A(E)* spectrum (**Fig 1.b**). However, the influence of increasing $E_U$ on the broadening of the absorption edge is rather minor.

In **Fig 1. c** we calculated $\delta\Delta\mu^{rad}$ for different $\sigma_g$ with increasing Urbach energy using **Eq. 3**. The range of values covers the range typically observed in the literature for CIGSe and other inorganic absorbers. Here it is crucial to define the reference band gap point ($E_g$) in **Eq. 3** to perform our calculations. In this particular case, the $\delta\Delta\mu^{rad}$ was calculated by considering the average band gap as a reference band gap point (i.e, $E_g = \overline{E_g} = 1.1 \, eV$ in **Eq. 3**). At first glance, it is evident that, similar to the error function model, an increase in $\sigma_g$ leads to an increase in $\delta\Delta\mu^{rad}$.

Moreover, Urbach tails contribute further to the $\delta\Delta\mu^{rad}$. As the $E_U$ rises, the $\delta\Delta\mu^{rad}$ also increases.

Based on our simulations, we can observe that band gap fluctuations cannot be evaluated similarly to Urbach tails. Although both parameters contribute to the broadening of *A(E)* and increase in $\delta\Delta\mu^{rad}$, the influence of Urbach tails is more pronounced in the energy ranges deep into the band gap (around 200 meV below $\overline{E_g}$). Therefore, in the experimental measurements, it is very important to use the energy ranges very deep into the band gap, i.e. very low α(E), to extract $E_U$ [46, 47]. Determining $E_U$ at higher energies closer to the band gap would be influenced mainly by $\sigma_g$ [47]. See also **Appendix D** for a comparison between the effect of Urbach tails and the effect of band gap fluctuations.

CIGSe absorbers are reported to have $E_U$ in the range of 11 to 20 meV [12, 20, 48, 49]. In **Fig 1. c** we show that within this range, the Urbach tails have minimal impact on the $\delta\Delta\mu^{rad}$. Additionally, In the case of $\sigma_g$ = 0 meV, we observe that at small $E_U$, the $\delta\Delta\mu^{rad}$ has negative values. This indicates that $\Delta\mu^{rad}$ is greater than the SQ limit. Although this might seem surprising at first, this effect is due to the reduced absorptance above the band gap with the square root increase in the absorption coefficient (α) which leads to a gradual increase in *A(E)* for energies higher than $E_g$ (See **Fig 2. a**) Consequently, this, reduces the integral of *A(E)* over $\Phi_{BB}(E)$ and reduces $F_0^{rad}$ to values lower than $F_0^{SQ}$, resulting in negative values for $\delta\Delta\mu^{rad}$(**Eq. 3**). In **Fig 2. a**, for the case of $\sigma_g$ = 0 meV and $E_U$ = 0 meV, we have illustrated the SQ-*A(E)* (i.e., step function) and *A(E)* extracted by considering the square root behavior of the absorption coefficient (α) above the band gap (SQRT model). Additionally, in **Fig 2. b** we compared the $F_0^{rad}$ values for each emission spectrum. Lower $F_0^{rad}$ means a higher quasi-Fermi level splitting is needed to balance the generation flux by the emission flux, i.e. negative $\delta\Delta\mu^{rad}$.

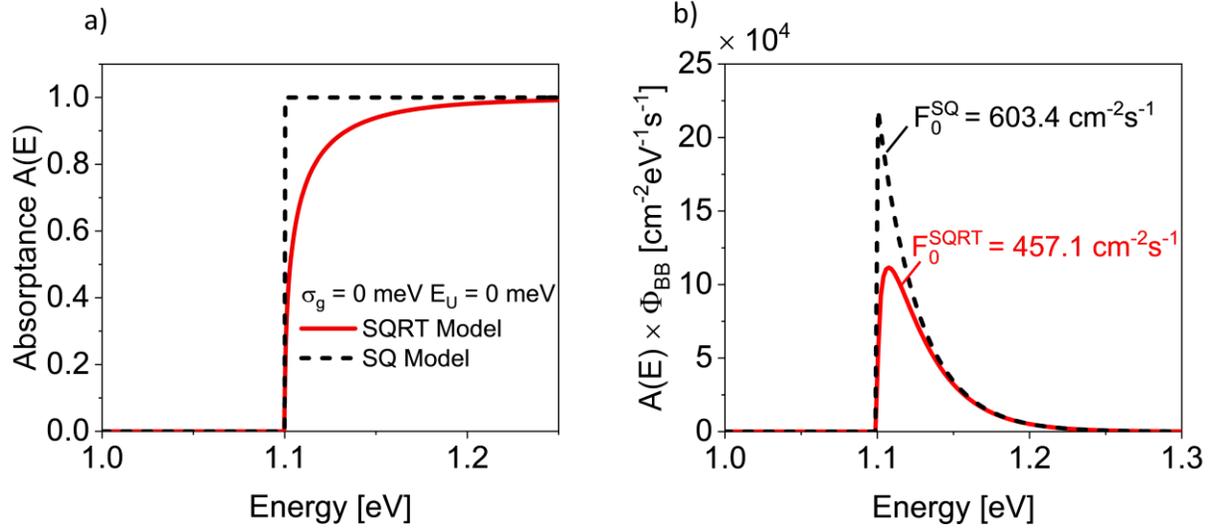

**Fig 2. a)** *A(E)* spectrum for an absorption coefficient with square root behavior above the band gap (SQRT) and for a step function (SQ-model). **b)** $A(E) \times \Phi_{BB}$ spectra for each *A(E)* spectra. $F_0^{rad}$ values for each *A(E)* are calculated by extracting the area under these curves (i.e., $F_0^{rad} = \int_0^\infty A(E)\Phi_{BB}(E)dE$ ). $F_0^{SQRT}$ and $F_0^{SQ}$ are $F_0^{rad}$ values obtained for SQRT and SQ model respectively (We consider an absorber thickness $d$ = 2μm.)

## 4. Measurements on CIGSe absorbers with and without band gap gradient:

### 4.1. *Δμ* losses in CIGSe absorbers deposited on glass

Up to this point, we have discussed the impact of band gap fluctuations and Urbach tails on the broadening of the absorptance edge. In this section, we will apply this analysis to experimentally measured *A(E)* spectra, and we quantify the *Δμ* and its losses. At the beginning, we investigate five distinct samples deposited on glass (soda lime glass SLG). These samples cannot be used to make solar cells, since they miss the back contact, but they allow us the independent determination of the *A(E)* using a photospectrometer. The name, composition as measured by Energy-Dispersive X-ray spectroscopy (EDX), and key properties of each sample are summarized in **Table 2**. Our study includes non-graded CIGSe samples (denoted as **NG**) as well as band gap-graded CIGSe samples (denoted as **BGG**). For selected samples we performed secondary ion mass spectrometry (SIMS) measurements to evaluate the presence of gradient (see **Fig S4** in **SI**).

Our analysis starts with the extraction of the *A(E)* spectrum for each sample. *A(E)* spectra are obtained by conducting transmittance and reflectance measurements on each absorber using a photospectrometer. From transmittance *T(E)* and reflectance *R(E)* spectra, the *A(E)* spectra can be calculated as *A(E) = 1 – R(E) – T(E)*. These measurements cover a wide range of photon energies. However, they are known to be unreliable for low absorptance values at the low energy slope (i,e near and below the band gap) [46]. On the other hand, PL measurements have high sensitivity for low absorptance values, i.e. at energies below the band gap energy [11, 46, 50]. Therefore, PL measurements are used to construct the low energy part of *A(E)* spectra.

According to generalized Planck's law of radiation [1] (**Eq. 1**), with knowledge of the PL flux spectrum ($\Phi_{PL}(E)$) and temperature of the absorber (here 296 K), we can obtain the non-absolute shape of *A(E)* using the following relationship [1, 50]:

$$A^{PL}(E) \propto \frac{\Phi_{PL}(E)}{\Phi_{BB}(E)} \qquad (Eq.19)$$

Here *A$^{PL}$(E)* denotes the *A(E)* extracted from PL measurements. It is non-absolute, since at this point, we do not know the $\Delta\mu$, yet, which according to **Eq. 1** enters exponentially into the proportionality factor in **Eq. 19**. However, in the energy ranges where both PL and photospectrometry measurements are reliable, the *A$^{PL}$(E)* can be aligned with the absorptance spectrum measured directly from transmittance and reflectance *A$^{dir}$(E)*. By re-scaling the *A$^{PL}$(E)* spectra to those obtained from the spectrometer, we can effectively construct the full *A(E)* spectrum across the entire energy range. Then we use the full range *A(E)* spectra to calculate the $\Delta\mu^{rad}$ and $\Delta\mu$ losses.

In **Fig 3**, the top row displays the *A(E)* spectra of samples A-NG, B-NG and C-BGG derived from both PL and transmittance/reflectance measurements. Additionally, the *A(E)* spectra of samples D-NG and E-BGG can be found in the SI (see **Fig S5** in **SI**).

**Table 2.** $\frac{Cu}{Ga+In}$ (CGI), $\frac{Ga}{Ga+In}$ (GGI) and presence of gradient for samples deposited on SLG. CGI and GGI are calculated independently from EDX data. NG denotes samples without gradient, BGG samples with band gap gradient.

| Sample Name | CGI | GGI | Gradient |
|---|---|---|---|
| A-NG | 0.90 | 0.20 | No |
| B-NG | 0.94 | 0.26 | No |
| C-BGG | 0.88 | 0.30 | Yes |
| D-NG | 0.88 | 0.25 | No |
| E-BGG | 0.94 | 0.20 | Yes |

To determine $\Delta\mu^{rad}$, the numerical value of the band gap energy is not needed, see (**Eq. 2**): for $\Delta\mu^{rad}$ calculations, it is sufficient to know the *A(E)* spectrum and the temperature of the absorber. However, to quantify $\Delta\mu$ losses, we need to calculate the SQ values, $F_{Gen}^{SQ}$ and $F_0^{SQ}$. Thus, it is necessary to know the band gap.

To determine the band gap energy, we apply a Gaussian fit to the first derivative of the *A(E)* spectra, and we take the maximum of this fit as an average band gap $\overline{E_g}$. We use $A^{PL}(E)$ spectra for this analysis, as the direct measurements $A^{dir}(E)$ tend to become unreliable near the absorption onset, and their derivatives of $A^{dir}(E)$ are often unsuitable for accurate band gap and broadening determination (see **Fig S6 SI**). It is important to mention that this $\overline{E_g}$ value is a reference point for our calculations. It is evident that the determination of $\overline{E_g}$ with 1 meV accuracy can be challenging. However, small errors in $\overline{E_g}$ determination would have compensating effects on the SQ limit and $\Delta\mu$ losses in a way that the final radiative limit ($\Delta\mu^{rad}$) would not change. This is due to the fact that $\Delta\mu^{rad}$ does not depend on the defined band gap according to **Eq. 2**.

In **Fig 3,** the bottom row illustrates the $\overline{E_g}$ and band gap distribution obtained from the first derivative of $A^{PL}(E)$. This approach is similar to the error function model (**Eq. 17**), where the inflection point of $A(E)$ yields $\overline{E_g}$. Additionally, the broadening of the Gaussian fit gives us a degree of broadening for the edge of $A(E)$. However, it is important to mention once again that the experimental absorptance edge has a more complicated function and depends on many factors as we discussed in the previous section. Still, the Gaussian fit describes the derivative of the absorptance spectrum surprisingly well. That means, that a Gaussian distribution of step-like SQ band gaps is a very suitable model to describe these absorptance edges. In some cases, we see that the experimental $A(E)$ derivative deviates from the Gaussian fit model. But even in these cases, valuable information can be extracted from the Gaussian fit, enabling a comparison between the samples.

For quantifying the $\Delta\mu$ and relative losses, in the first step we quantify $qV_{OC}^{SQ}$ (i.e., $qV_{OC}^{SQ} = kT \, Ln\left(\frac{F_{SC}^{SQ}}{F_0^{SQ}}\right)$) using the obtained band gap from inflection point of $A(E)$. It is worth mentioning that all our PL measurements are performed at the room temperature (here 296 K), thus, our calculations are performed by considering T=296 K.

Our next step is quantifying the radiative $\Delta\mu$ loss ($\delta\Delta\mu^{rad}$) and generation $\Delta\mu$ loss ($\delta\Delta\mu^{Gen}$) with using **Eq. 3** and **Eq. 4**. The $\delta\Delta\mu^{rad}$ primarily depends on the broadening of the $A(E)$, for all our samples we calculated ($\delta\Delta\mu^{rad}$), the values are summarized in **Table 3**. As expected, with an increase in the σ$_A$ (i.e., *A(E)* edge broadening), $\delta\Delta\mu^{rad}$ increases (**Fig 4. a**). On the other hand, the $\delta\Delta\mu^{Gen}$ is mainly due to the incomplete absorption. In our samples, we observe that *A(E)* nearly plateaus to values lower than 1 at energies well above $\overline{E_g}$ (see **Fig 3. a-c**), and all samples plateau to similar values. Consequently, the $\delta\Delta\mu^{Gen}$ for these samples have similar values around 3 meV to 4 meV (see **Table 3**)

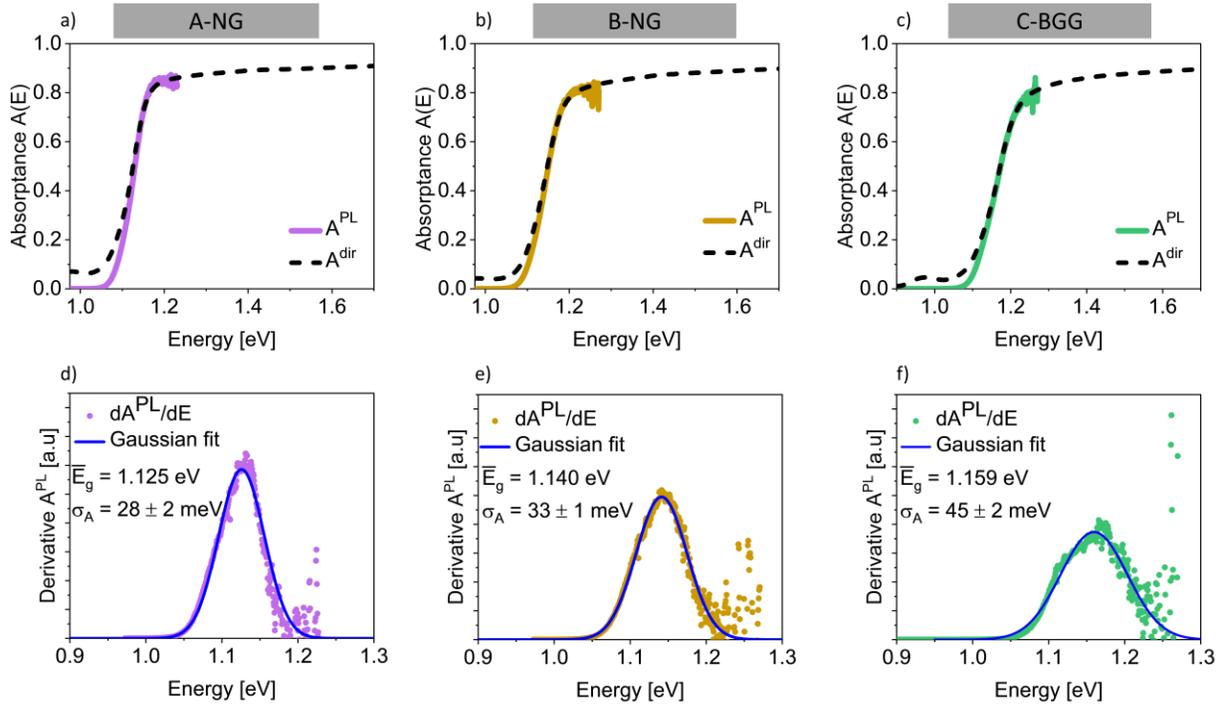

**Fig 3. a)**, **b)** and **c)**, absorptance *A(E)* spectra for CIGSe samples A-NG, B-NG and C-BGG respectively. $A^{dir}$ is the (absolute) absorptance spectrum measured by photospectrometry and $A^{PL}$ is the absorptance extracted from PL and scaled in the overlapping energy range to $A^{dir}$. **d)**, **e)** and **f)** Gaussian fit to the first derivative of the absorptance spectrum of samples A-NG, B-NG and C-BGG. To improve visualization, the 1$^{st}$ derivative here is extracted from the smoothed absorptance curves, as detailed in the **experimental section**. Different fitting ranges were used to extract σ$_A$, and the average value is reported here with an error range of approximately 1 meV to 2 meV.

For these samples, we observe that both, $\sigma_A$ and $\delta\Delta\mu^{rad}$, increase with the introduction of a Ga gradient (**Fig 4. a**). In the subsequent sections, we will discuss in more detail the effect of band gap gradient and compositional inhomogeneities on the broadening of the absorption edge.

Finally, in the last step, we determine the $Y_{PL}$ for each sample and calculate the $\delta\Delta\mu^{nr}$ using **Eq. 5**. It is crucial to note that for CIGSe absorbers, extracting the $Y_{PL}$ under 1 sun illumination condition is essential. The presence of metastable defects results in non-constant $Y_{PL}$ under different illumination intensities [27, 51, 52]. In **Fig S7** in the SI, we illustrate $Y_{PL}$ for these samples with increasing laser intensity: the $Y_{PL}$ in CIGSe samples increases with increasing incident light intensity. In all our calculations in the previous sections, we consider illumination under 1 sun incident intensity. Thus, for our calculations, the $\delta\Delta\mu^{nr}$ values are all extracted from measurements under 1 sun incident intensity.

From the complete absorptance spectrum and the $Y_{PL}$ we can determine $\Delta\mu^{rad}$ and all $\Delta\mu$ losses to extract the final $\Delta\mu$ value, according to **Eq. 6**. In **Fig 5. b,** we show the contribution of each loss mechanism and the resulting $\Delta\mu$ of our samples, the values for each loss and final $\Delta\mu$ values are summarized in **Table 3**.

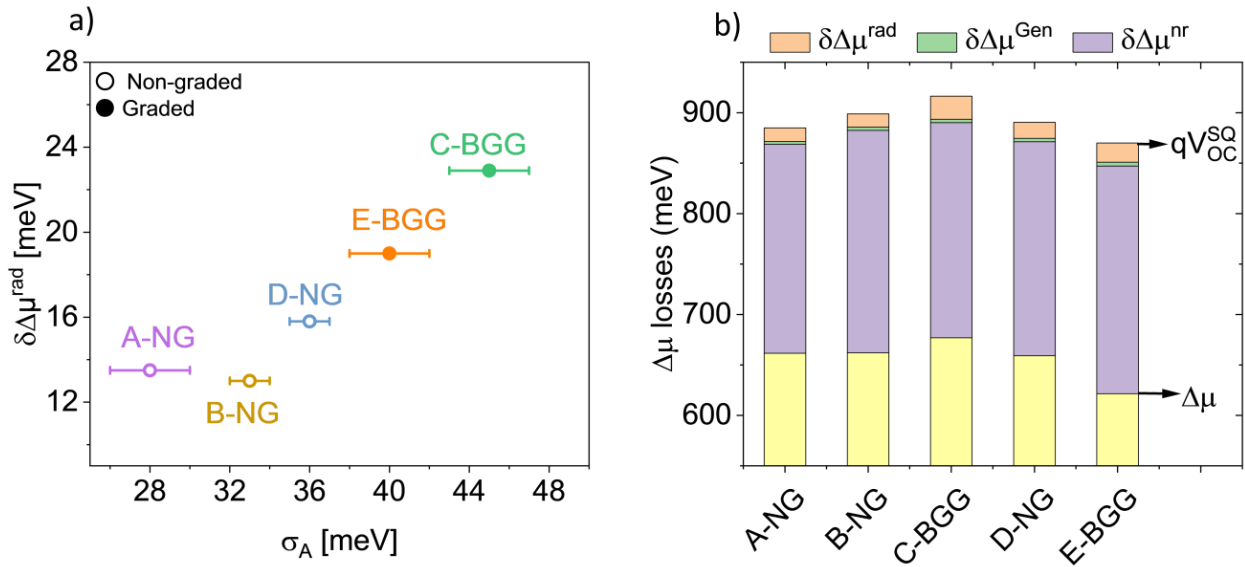

**Fig 4. a)** Increase in $\delta\Delta\mu^{rad}$ with increase in $\sigma_A$, different fitting ranges were applied to extract $\sigma_A$ from the Gaussian fit of $\frac{dA(E)}{dE}$. The average value of $\sigma_A$ is represented here, and the error bars denote the standard deviation. **b)** $\Delta\mu$ losses calculated for the samples deposited on SLG, the respective values are summarized in **Table 3**.

Moreover, the $\Delta\mu$ can also independently be extracted directly from Planck's generalized law [3, 4]. Using **Eq. 1** and Boltzmann approximation (i.e., $E - \Delta\mu \gg k_B T$) we can write it into the following form [3, 4].

$$ln\left(\frac{\Phi_{PL} h^3 c^2}{2\pi A(E) E^2}\right) = -\frac{E - \Delta\mu}{k_B T} \quad (Eq.\,20)$$

Here, the $\Delta\mu$ can be determined from the vertical axis intercept of a linear fit to the high-energy wing, while the measurement temperature can be determined from the slope of the fit [4, 20].

In our study, we have already determined the *A(E)* spectra at wide energy ranges from transmittance and reflection measurements. However, in many cases, a simple approximation for *A(E)* can be used to extract the *Δμ*. In the literature, often the approximation of *A(E)=1* for high enough energies is used [3, 4, 36, 53]. In our previous reports, we have already demonstrated that this assumption introduces errors of 10 meV to 20 meV [4]. In the cases where *A(E)* of the samples is constant at relatively high energy ranges above the band gap, assuming a constant value for A(E) (here *A(E)=1*) would result in the temperature extracted from this fit being the same as the measurement temperature (here ~296 K) [4]. However, when the *A(E)* spectra of the sample continues to increase and does not plateau, assuming constant values for *A(E)* (here *A(E)=1*), resulting in extracted temperatures higher than the measurement temperature and unrealistically small *Δμ* values [4, 11, 20, 46]. Nevertheless, by fixing the temperature to the measurement temperature, (i.e, fixing the slope), we enforce the *A(E)=1* condition and enhance the accuracy of our extracted *Δμ* values [3, 4, 11, 20, 27].

In **Fig 5. a**, we performed a linear fit to the high-energy slope of **Eq. 20** to determine the $\Delta\mu$ by considering both assumptions with $A(E) = A^{dir}(E)$ and $A(E) = 1$. (Here $A^{dir}(E)$ is directly measured absorptance from the spectrometer).

In our samples, when we consider $A(E) = A^{dir}(E)$ in **Eq. 20**. The free-slope fitted temperature is in the range of 296 ±2 K , which aligns closely with our measured temperature of 296 K (see **SI note 9**). Notably, this extracted temperature from linear fit is sensitive to the fitting range and local slope, which can be affected by the noise level in the measurements. We tested various fitting ranges, and the free-slope fit yielded a temperature with an error of ± 2 K. In **SI note 9**, we discuss in detail the effects of local slope fluctuations and fitting ranges on the extracted temperature from linear fit slope. In **Fig 5**, to improve consistency across all $\Delta\mu$ extractions, we applied a fixed slope for the linear fit in all cases, setting the temperature at the measured temperature of 296 K. For comparison, the assumption of $A(E) = 1$ was also employed and $\Delta\mu$ was obtained under this assumption by fixing the temperature slope (i.e, fixing temperature) of linear fit.

In **Fig 5. b,** we have summarized the $\Delta\mu$ values extracted from different methods, including calculations from losses (**Eq. 6**), using linear fit with known $A^{dir}(E)$ and linear fit with assuming $A(E)=1$.

We note that the assumption $A(E) =1$ underestimated the $\Delta\mu$ by 5 meV to 7 meV. This observation aligns well with our previous studies, where we addressed the limitations of this assumption [4]. On the other hand, using the measured $A^{dir}(E)$ provides more reliable results that closely match the calculated values from losses.

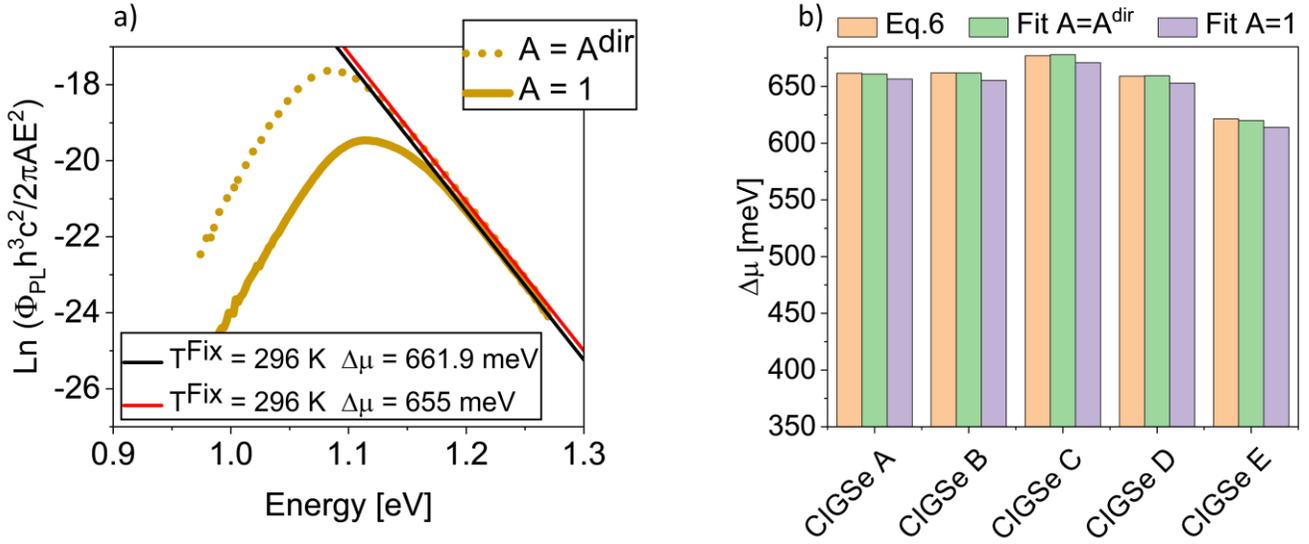

**Fig 5. a)** Linear fit to the high energy slope of modified Planck's generalized law (***Eq. 20***) to extract the *Δμ*. Here assumption of *A(E) = 1* and *A(E) = $A^{dir}$(E)* was employed and in both cases the *Δμ* was extracted by fixing temperature to 296 K (i.e, fixing slope). More details regarding the fit temperature can be found in **SI note 9**. **b)** Corresponding *Δμ* values extracted from calculation based on losses (using ***Eq. 6***) and linear fit to ***Eq. 20*** with *A(E) =1* and *A(E)=$A^{dir}$(E)* assumption (The *Δμ* values from each method are summarized in **Supplementary Table 1** in **SI**).

**Table 3** $\Delta\mu$ losses with respect to SQ limit, the $qV_{OC}^{SQ}$ and all values are calculated by considering the measurement temperature T=296 K. The final $\Delta\mu$ value is extracted from ***Eq. 6***.

| Sample Name | $\sigma_A$ (meV) | $\overline{E_g}$ (eV) | $qV_{OC}^{SQ}$ (meV) | $\delta\Delta\mu^{rad}$ (meV) | $\delta\Delta\mu^{Gen}$ (meV) | $\Delta\mu^{rad}$ (meV) | $\delta\Delta\mu^{nr}$ (meV) | $\Delta\mu$ (meV) |
|---|---|---|---|---|---|---|---|---|
| A-NG | 28 ± 2 | 1.125 | 885.0 | 13.5 | 2.8 | 869.7 | 207.1 | 661.6 |
| B-NG | 33 ± 1 | 1.140 | 898.8 | 13.0 | 3.3 | 882.5 | 220.5 | 662.0 |
| C-BGG | 45 ± 2 | 1.159 | 916.4 | 22.9 | 3.3 | 890.2 | 213.2 | 677.0 |
| D-NG | 36 ± 1 | 1.131 | 890.5 | 15.8 | 3.4 | 871.3 | 212.2 | 659.1 |
| E-BGG | 40 ± 2 | 1.109 | 869.9 | 19.0 | 3.8 | 847.1 | 225.6 | 621.5 |

The agreement between the $\Delta\mu$ determined from the combined effects of radiative, generation, and non-radiative losses, and the fit to Planck's generalized law, is a key observation, and indicates that the model for the various $\Delta\mu$ or $V_{OC}$ losses is complete. It is important to note that these $\Delta\mu$ values are extracted from a single point on the absorber. The aim of this analysis is to demonstrate that the $\Delta\mu$ loss analysis is complete, therefore we have performed this analysis for single point on the absorber to avoid any uncertainties in the $Y_{PL}$ inhomogeneities across the absorber [7]. Therefore, the $\Delta\mu$ values here were presented without statistical error bars.

## 4.2 Comparison with Solar Cells

Up to this point, our analysis has been based on optical absorptance and photoluminescence measurements. For solar cells, which are generally deposited on an opaque Mo back contact this approach is not possible. However, as highlighted in **Table 1**, similar calculations based on the reciprocity principle can be carried out using quantum efficiency of the short circuit current $QE(E)$ spectra to extract relative $V_{OC}$ losses of a photovoltaic device [7, 19].

In this section, we extend our analysis to conduct a $V_{OC}$ loss assessment for three CIGSe solar cells: F-BGG sample, which incorporates a band gap gradient across the absorber, and G-NG and H-NG samples both exhibiting a uniform band gap throughout their thickness. Notably, the H-NG sample is pure CISe sample without any Ga. On each absorber film we prepare 5 or 8 solar cells with areas of ~ 0.25 cm$^2$. In the following we discuss the properties of the best solar cell on each absorber. The performances of all investigated cells are given in **SI note 12**.

**Fig 6. a** illustrates the J-V characteristic of the solar cell with the highest efficiency on each absorber, and corresponding device parameters along with the composition of the absorbers, are summarized in **Table 4**.

**Table 4**. Key parameters of solar cells investigated in this study. Including $\frac{Cu}{Ga+In}$ (CGI), $\frac{Ga}{Ga+In}$ (GGI) values along with the photovoltaic parameters of the champion solar cell on each absorber. The J-V characteristics are shown in **Fig 6. a**.

| Sample name | Gradient | CGI | GGI | V$_{OC}$ (mV) | J$_{SC}$ (mA/cm$^2$) | FF | Eff % |
|---|---|---|---|---|---|---|---|
| F-BGG | Yes | 0.93 | 0.25 | 648 | 35.5 (35.1*) | 76.7 | 17.6 |
| G-NG | No | 0.94 | 0.27 | 608 | 33.7 (33.4*) | 76.1 | 15.6 |
| H-NG | No | 0.99 | 0 | 474 | 40.8 (40.4*) | 71.1 | 13.7 |

\* Extracted from integration of QE(E) over AM 1.5 g.

**Fig 6. b** depicts the QE spectra of the samples. Similar as $A^{dir}(E)$ spectra, the directly measured $QE^{dir}(E)$ spectra level off at low values below the absorption edge, because of the limited sensitivity of the equipment. This effect is commonly known in the literature [13, 54]. As illustrated in **Table 1**, the relationship between the *QE(E)* spectra and the EL spectra $\Phi_{EL}(E)$ are the same as between absorptance and the PL spectra $\Phi_{PL}(E)$ (see ***Eq. 1*** and ***Eq. 7***). Therefore, to construct the low energy decay of the *QE(E)* spectra, additional EL measurements were carried out. The reciprocity relationship between EL emission and *QE(E)* allows us to construct the low energy decay of *QE(E)* spectra from EL measurements (referred as $QE^{EL}$) (see ***Eq. 7***) [7, 13, 19, 54, 55]. Similar, to the PL spectra analysis, we extract a non-absolute *QE(E)* spectrum from the EL spectra and adjust it in the overlapping energy range to the $QE^{dir}$ spectrum. Furthermore, we use the inflection point of the *QE(E)* spectra to evaluate the absorption edge broadening and $\overline{E_g}$ value (**Fig 6. c**). Notably, in contrast to optical measurements, here both $QE^{dir}$ and $QE^{EL}$ can be used for this purpose. (see **Fig S11 a and c** in the SI). Both spectra give the same broadening parameter (within error of ± 2 meV). However, for simplicity here we just used the $QE^{dir}$ measurements for band gap and broadening extraction. In **Fig 6. c** we extracted the broadening of the SQ band gap distribution $\sigma_{QE}$ from $QE^{dir}$ measurements by fitting again a Gaussian to the first derivative of the $QE^{dir}$ spectrum.

As we mentioned previously, the *QE(E)* and *A(E)* are linked through the collection function (See **SI note 2**) [33, 34]. At energy ranges near the band gap energy and below it (i.e., with small absorption coefficient values), the collection function can be considered as a constant factor that scales *A(E)* to *QE(E)* values [34]. Thus, the *A(E)* and *QE(E)* spectra are expected to have the same shape in this region. Therefore, the inflection point of *QE(E)* represents the onset of *A(E)*. Notably, in our samples, the EL and PL spectra have the same shape, and the *A(E)* and *QE(E)* spectra are essentially identical (see **Fig S11. a and c** in **SI**).

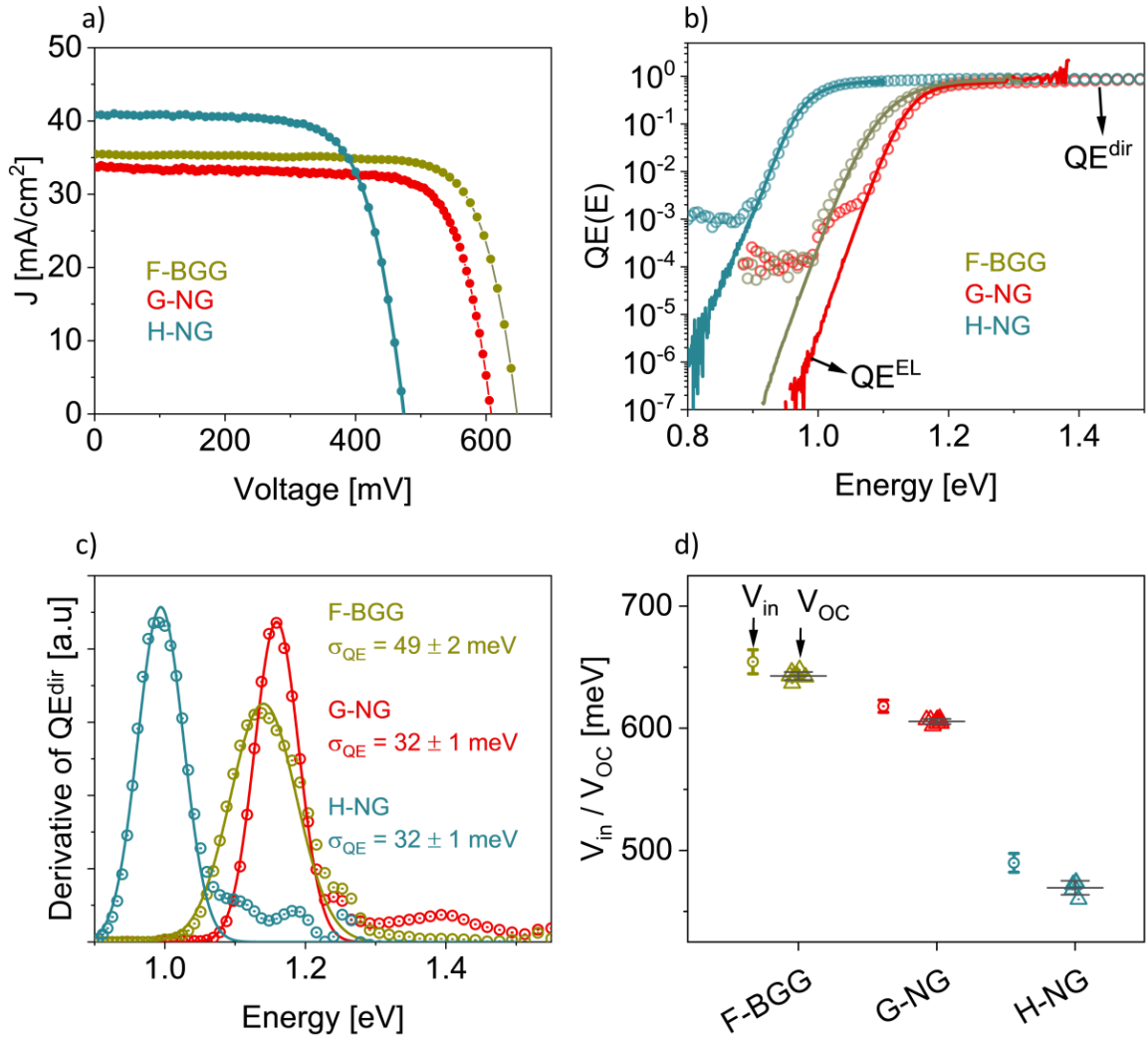

**Fig 6** Each sample is represented by a different color **a)** J-V curve of the solar cells with highest efficiency on each absorber **b)** *QE(E)* spectra extracted for samples F-BGG, G-NG and H-NG with combination of direct measurements (*QE$^{dir}$*) and EL measurements (*QE$^{EL}$*). **c)** First derivative of *QE$^{dir}$* along with the $\sigma_{QE}$ extracted for each sample. The $\sigma_{QE}$ here is extracted from gaussian fit to first derivative of *QE$^{dir}$*. Different fitting ranges were employed, and average value is reported here with the errors representing the standard deviation. **d)** Comparison between *V$_{in}$* and measured *V$_{OC}$* for each sample. The uncertainty in V$_{in}$ is due to errors in $\delta V_{OC}^{nr}$ determination. (More details of EL measurements are explained in the experimental section.)

To determine the relative voltage losses, we start our calculations by obtaining the $V_{OC}^{SQ}$ from the extracted band gap. Subsequently, we quantify the radiative voltage loss $\delta V_{OC}^{rad}$ and short circuit voltage loss $\delta V_{OC}^{SC}$ using **Eq. 9** and **Eq. 10** provided in **Table 1**, and finally, we determine the radiative voltage limit ($V_{OC}^{rad}$) and non-radiative voltage losses ($\delta V_{OC}^{nr}$) for each sample using $Y_{EL}$ (see **Eq. 14**)

In **Fig 6. c** we observe that sample F-BGG with graded band gap profile has a broader $QE(E)$ onset than the other two samples. Consequently, this broadening leads to increased radiative voltage loss ($\delta V_{OC}^{rad}$). Our calculations indicate that F-BGG sample with band gap gradient shows ~16 meV higher radiative $\delta V_{OC}^{rad}$ compared to G-NG and H-NG samples. The relative voltage losses for these samples are summarized in **Table 5**.

**Table 5**. $V_{OC}$ losses with respect to SQ limit, all calculations were performed by considering measurement temperature of T = 296 K. Overall the $\delta V_{OC}^{nr}$ were measured from different cells. The average value for $\delta V_{OC}^{nr}$ is represented here with errors indicating the standard deviation. The errors induced in $V_{in}$ is due to same uncertainty in $\delta V_{OC}^{nr}$.

| Sample Name | $\sigma_{QE}$ (meV) | $\overline{E_g}$ (eV) | $V_{OC}^{SQ}$ (mV) | $\delta V_{OC}^{rad}$ (mV) | $\delta V_{OC}^{SC}$ (mV) | $V_{OC}^{rad}$ (mV) | $\delta V_{OC}^{nr}$ (mV) | $V_{in}$ (mV) |
|---|---|---|---|---|---|---|---|---|
| F-BGG | 49 ± 2 | 1.140 | 898.8 | 25.8 | 5.1 | 867.9 | 213.4 ± 9.8 | 654.5 ± 9.8 |
| G-NG | 32 ± 1 | 1.159 | 916.4 | 10.0 | 5.8 | 900.6 | 282.5 ± 5.0 | 618.1 ± 5.0 |
| H-NG | 32 ± 1 | 0.994 | 762.8 | 10.0 | 4.6 | 748.2 | 258.4 ± 7.6 | 489.8 ± 7.6 |

In the next step, we extract the $\delta V_{OC}^{SC}$, analogous to $\delta\Delta\mu^{Gen}$. For $\delta V_{OC}^{SC}$, incomplete absorption and collection losses can both contribute. An increase in the collection losses can reduce the *QE(E)*, lower the *J<sub>SC</sub>*, and increase the $\delta V_{OC}^{SC}$. In our samples, the $\delta V_{OC}^{SC}$ is around 5 meV to 6 meV. (see **Table 5**) and its overall contribution to voltage losses is still almost negligible, but somewhat larger than the pure generation losses extracted from absorptance (compared to **Table 3**). Additionally, by performing absolute EL measurement under an injected current of *J<sub>SC</sub>* (equivalent to 1 sun condition) on a complete device we can obtain *Y<sub>EL</sub>* and non-radiative voltage loss (**Eq. 14**) and subsequently we extract the internal voltage *V<sub>in</sub>* using **Eq. 12** in **Table 1**. Here, *V<sub>in</sub>* is completely analogous to *Δμ*, and it represents the upper limit for *V<sub>OC</sub>*. In fact, in the absence of interface voltage losses or voltage dependent charge collection losses, *V<sub>in</sub>* and *V<sub>OC</sub>* should have the same values [3, 56, 57]. In **Fig 6.d,** we illustrate the *V<sub>in</sub>* and the experimentally measured *V<sub>OC</sub>* of our devices. Our analysis in **Fig 6.d** indicates that the *V<sub>in</sub>* exhibits slightly higher values (approximately 10 meV) compared to *V<sub>OC</sub>*. This discrepancy can be attributed to additional non-radiative recombination at the interface or voltage dependent charge collection losses which can reduce *V<sub>OC</sub>* below *V<sub>in</sub>* [56, 57].

In **Table 5,** we observe that the primary loss mechanism remains $\delta V_{OC}^{nr}$, attributed to non-radiative recombination channels. Notably, the non-graded G-NG and H-NG exhibits a higher $\delta V_{OC}^{nr}$ compared to the F-BGG sample, which incorporates a Ga gradient. This observation aligns with expectations, as the presence of the Ga gradient effectively passivates the back surface recombination at the CIGSe/Mo interface [58]. However, we do see that graded samples, compared to non-graded absorbers can suffer from ~16 meV higher radiative losses which in the state-of-the-art technology can be considerable. For a more comprehensive analysis, section 5 includes a broadening analysis conducted on 19 samples from our laboratory. Additionally, we conduct a meta-analysis of published data from various studies to compare the absorption edge broadening across different thin-film technologies.

**4.3 Microscopic analysis**

To shed light on the origin of the broadening, we performed cathodoluminescence (CL) measurements on CIGSe samples F-BGG, G-NG and H-NG, using a dedicated scanning electron microscope (SEM).

To reveal the microscopic origin of the broadening, we carried out CL on both cross-sections and surfaces of the absorbers. The SE images and panchromatic CL intensity maps can be found in **SI note 13**. The CL emission peak is related to the near band edge recombination. Therefore, the maximum CL emission peak maps allow us to access the band gap variations from grain to grain and within a single grain. **Fig 7. a** illustrates the cross-section CL map of the F-BGG sample with the presence of Ga gradient. To investigate the band gap variation across the thickness, we performed a line scan through the depth of the absorber. We observe a continuous blue shift in the CL peak position as we move from the surface to deeper regions (**Fig 7. b** and **c**). These results are in line with secondary ion mass spectrometry (SIMS) measurements (see **SI note 4**), which showed a steady increase in Ga concentration throughout the depth. However, a smooth increase of the Ga concentration in SIMS does not necessarily indicate a smooth band gap gradient on a microscopic scale [59]. Indeed, the shift of the CL maximum energy is not as smooth as the SIMS profile. The heat map in **Fig 7. c** shows a jump in the maximum energy around 0.35 µm depth. In contrast, for the two other samples, without compositional gradient, we observed a homogeneous band gap distribution across their thickness. (**see Fig S12** and **Fig S13 in SI**)

To also investigate lateral band gap fluctuations, we performed CL measurements on the surface of the bare absorbers. The resulting data, as shown in **Fig 7. d-f**, represent the fitted energy of the CL emission peak at each point across the surface. For each absorber, we calculated the CL standard deviation ($\sigma_{CL}$) from the distribution of the fitted CL emission energies in each pixel, which indicate the fluctuations from the average CL peak emission energy. The $\sigma_{CL}$ serves as an indicator of lateral band gap fluctuations, with higher values reflecting greater inhomogeneities in the band gap distribution across the surface of the absorbers. $\sigma_{CL}$ for each absorber is summarized in **Fig 7. d-f**. This approach provides a useful insights about the lateral variations in the band gap within the material.

We observed that the H-NG sample, which does not contain Ga, exhibits the smallest band gap fluctuations, indicated by the lowest value of $\sigma_{CL}$. On the other hand, the introduction of Ga results in an increase in $\sigma_{CL}$ to 8.5 meV for the G-NG sample. Additionally, the F-BGG sample, which features a Ga gradient, shows the highest $\sigma_{CL}$, indicating significantly stronger lateral band gap fluctuations compared to the two other samples.

Our CL observations provide insight into why the F-BGG sample with a Ga gradient exhibits more significant broadening in the absorption onset compared to the other two samples. Firstly, the effect of the depth gradient is crucial, as the absorption onset is determined by the position of the minimum band gap. In this particular sample, the minimum band gap position is confined to a very thin region near the surface of the absorber. This reduced thickness reduces the absorptance near the minimum band gap, resulting in a wider absorption edge.

Secondly, this sample also displays stronger lateral band gap fluctuations. As discussed earlier, samples with more pronounced lateral band gap variations show increased broadening in the absorption onset. It seems that the introduction of a gradient along the sample depth also increases the lateral fluctuations. However, it cannot be determined at the moment, whether this observation is specific to this sample or is general feature of Ga graded CIGSe samples.

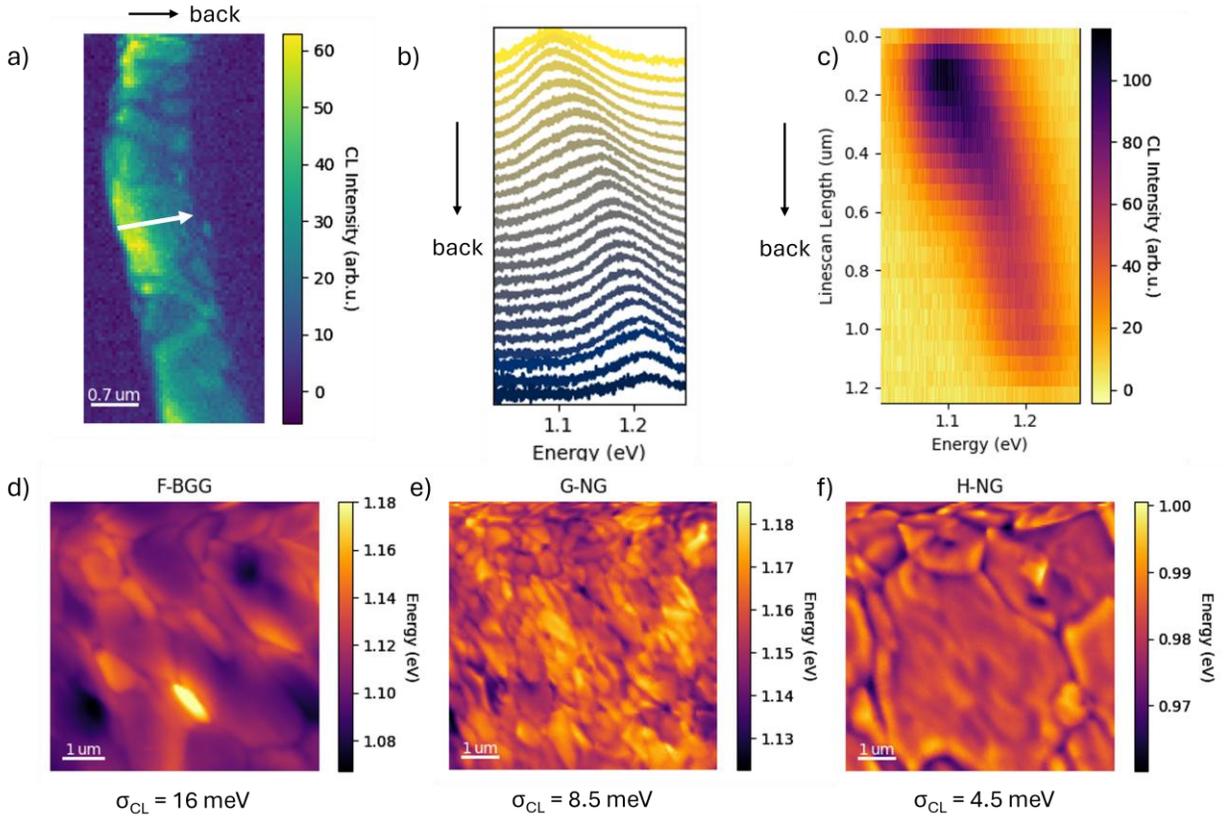

**Fig 7 a)** A panchromatic cross-section CL intensity map for the F-BGG sample, with the line scanned area shpwn with white arrow. The distortion in the CL map is likely due to charging effects. **b)** The evolution of the normalized CL spectra for the F-BGG sample as the line scan progresses deeper into the material. **c)** A "heat map" illustrating the blueshift of the CL peak along the line scan for sample F-BGG. **d), e),** and **f)** Color maps displaying the fitted energy of the CL emission spectra at each point for samples F-BGG, G-NG, and H-NG, respectively, along with their statistical standard deviations, which represent the variation from the mean CL emission energy value.

## 5. Discussion and meta-analysis

So far, we have experimentally quantified the relative $\Delta\mu$ or $V_{OC}$ losses of our CIGSe samples. By definition, radiative losses ($\delta V_{OC}^{rad}$ and $\delta\Delta\mu^{rad}$) increase with the increase in the broadening of the $A(E)$ or $QE(E)$ onset. Moreover, we have observed that a band gap gradient can contribute significantly to this broadening. To check the significance of these findings, we have investigate 19 samples from our laboratory (including samples in this study) to identify the primary cause of absorption edge broadening in our CIGSe absorbers.

We would like to remind ourselves that the broadening of the absorption edge can be related to many factors such as Cu content (CGI), Ga content (GGI) or Urbach tails that are often interconnected. For instance, lower CGI and higher GGI values can increase the magnitude of electrostatic potential fluctuation which can subsequently increase the Urbach energy and broadening parameter [21, 60].

**Fig 8** shows the broadening parameter values ($\sigma_A/\sigma_{QE}$) for different CGI, GGI and Urbach energies for both graded and non-graded samples from different processes. The Urbach energies are obtained from the absorption coefficient ($\alpha$) extracted from PL or EL measurements, where we extract the Urbach energies from fitting the low energy decay part of the $\alpha$ spectrum (see **SI note 11**). In **Fig 8** we have not separated $\sigma_A$ and $\sigma_{QE}$, since at the range where we extract the broadening parameter, the EL and PL spectra have similar shape. Therefore, the *A(E)* has identical shape as *QE(E)* for our samples (see **Fig S11**. **a** and **Fig S11**. **c** in SI). Thus, we used these parameters interchangeably to compare our samples.

The Urbach energy for these samples ranges from 11 meV to 18 meV, which aligns well with previous reports, where Urbach energy was derived from PL measurements [11, 20, 21, 48, 61]. It is important to note that PL or EL measurements allow us to measure ultralow values of the absorption coefficient deep into the band gap. As we showed previously, with EL and PL measurements, we can measure *QE(E)* and *A(E)* deep into the band gap, this capability enables us to determine the exponential decay of $\alpha$ deep into the band gap and obtain reliable Urbach energies [11] (see **Fig S11. b** in **SI**).

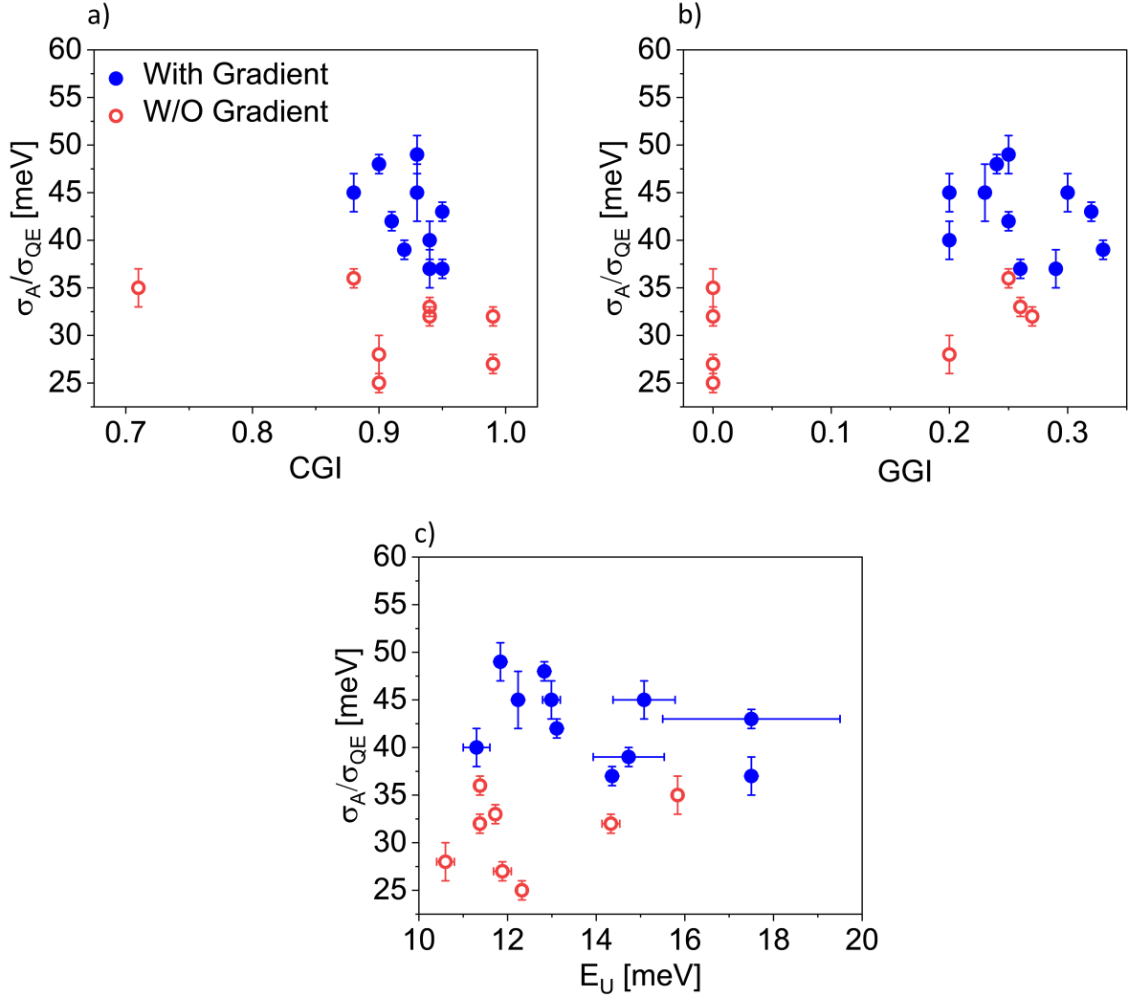

**Fig 8**. Broadening parameter ($\sigma_A/\sigma_{QE}$) of the absorption edge plotted versus **a)** CGI, **b)** GGI and **c)** Urbach energies for 19 CIGSe absorbers from different processes from our laboratory.

All samples in this study exhibit a Cu-poor composition (i.e., CGI < 1), with CGI values ranging from 0.7 to 0.99 and GGI values between 0 and 0.35.

Initially, in **Fig 8. a** and **b**, no clear trend is observed between the variation in CGI and GGI and changes in the $\sigma_A/\sigma_{QE}$. This does not mean that the absorption edge broadening does not depend on Cu or Ga content. But in this set of samples, different parameters were varied at once. It is important to note that this analysis is limited to these 19 samples. A more detailed statistical analysis would be needed to further investigate the effect of CGI or GGI on broadening, considering samples with constant CGI and varying GGI and vice versa, but here we would like to concentrate on the effect of band gap gradient. Moreover, we also see that the Urbach energy is almost uncorrelated to the overall broadening parameter of the samples. In both graded and non-graded samples, an increase in Urbach energies does not necessarily lead to an increase in the broadening values. We observe some samples with higher Urbach energies that have similar broadening compared to those with lower Urbach energies.

On the other hand, we observe that in all cases the samples which incorporate Ga gradient show larger broadening values. This analysis indicates that the primary factor contributing to the broadening of the absorption edge in our CIGSe absorbers beyond 35 meV is the presence of the band gap gradient throughout the thickness of the absorber.

To further investigate and understand the influence of the gradient on the broadening, we conducted a meta-analysis and reviewed published reports to analyze the relationship between broadening and gradient profiles. The results of this analysis are summarized in **Fig 9**. Additionally, data from other inorganic semiconductors, such as GaAs, CdTe, CZTS, and Si solar cells, are also included for comparison.

In **Fig 9**., the CIGSe absorbers are classified into three distinct gradient profiles: non-graded (**NG**) samples, characterized by a uniform band gap distribution throughout the depth; hockey-stick graded (**HSG**) samples, which exhibit a constant band gap over most of the depth with a sharp increase near the back contact; and band gap graded (**BGG**) samples, where the band gap varies significantly over the thickness. Importantly, in both NG and HSG samples, the effective absorption thickness extends significantly across the absorber depth, while in BGG samples, this effective thickness is confined to a narrow region close to the surface.

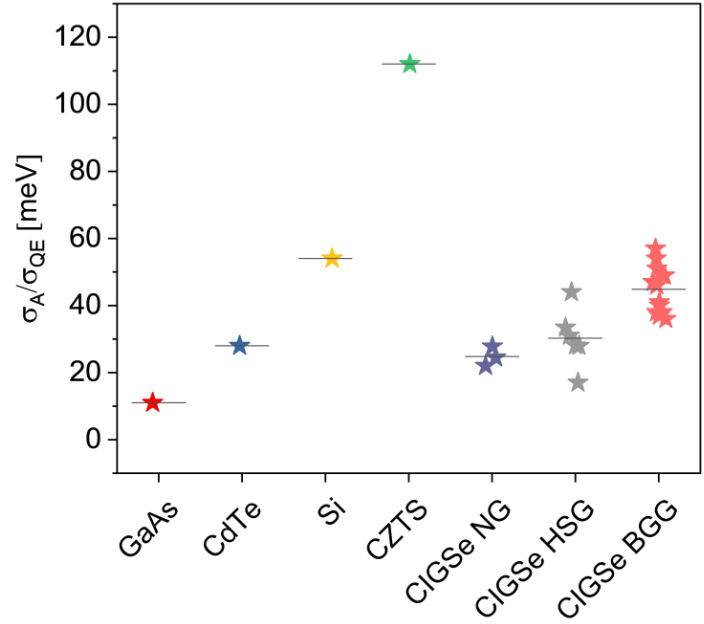

**Fig 9**. Meta-analysis presenting the comparison of broadening values ($\sigma_A/\sigma_{QE}$) from various published studies for different solar cell absorbers, including GaAs [62], CdTe [63], Si [64], CZTS [65], CIGSe absorbers with a hockey stick gradient (HSG) [12, 38, 58, 66], non-graded (NG) CIGSe absorbers [58, 67], and band gap graded (BGG) absorbers [12, 20, 48, 68, 69]. For more details regarding the meta-analysis see the experimental and methods.

In **Fig 9,** we observe that NG and HSG CIGSe absorbers generally exhibit lower broadening values compared to BGG samples. This trend aligns with our findings. Among the compared technologies, GaAs stands out with the lowest broadening value of approximately 11 meV, which is 6 meV lower than the lowest reported $\sigma_A/\sigma_{QE}$ value for CIGSe. CdTe technology shows broadening values comparable to those of NG and HSG CIGSe samples. Meanwhile, BGG CIGSe samples exhibit broadening as high as those observed in Si solar cells with indirect band gap. The highest value of broadening is observed in CZTS absorbers, which are well known for their strong band gap fluctuations and significant density of states within the band gap (i.e., pronounced Urbach tails) [42, 70].

Overall, investigation of our samples combined with the meta-analysis of published results, highlights that the band gap gradient is significant factor contributing to the broadening of *QE(E)/A(E)* edge. We can conclude that to mitigate the radiative losses ($\delta V_{OC}^{rad}$ and $\delta \Delta \mu^{rad}$), CIGSe absorber with uniform band gap profile should be prepared. However, in this case the passivation of the back contact region is essential. The HSG band gap profile has shown to be effective method to passivate the back surface recombination. Notably, the recent record efficiency CIGSe solar cell was prepared using HSG band gap profile [38]. Moreover, in recent years a hole transport layer and selective back contact were also developed for CIGSe absorbers and can effectively mitigate the back surface recombination [58].

## 6. Conclusion

In this manuscript, we employed the generalized detailed balance model to investigate the relative *Δμ* and voltage losses for solar cell absorbers with and without a band gap gradient. Our primary focus was on the radiative losses ($\delta \Delta \mu^{rad}$ and $\delta V_{OC}^{rad}$). Initially, we simulated the effect of band gap fluctuations and sub-band gap states (i.e., Urbach tails) on the radiative *Δμ* losses. In our simulations, we considered Gaussian distribution of step-function band gaps and realistic absorption coefficient of material with square root increase for energies above the band gap, as well as Urbach like decays for energies below the band gap energy. Furthermore, we employed a generalized detailed balance model to experimentally measured absorptance spectra. The *A(E)* is obtained from a combination of photospectrometry and photoluminescence data. Our experimental results are in line with our simulation, where we observe that $\delta \Delta \mu^{rad}$ increases with the increase in the broadening of *A(E)* ($\sigma_A$). Additionally using analogy between *QE(E)* and *A(E)* measurement, we quantified *V_{OC}* losses for CIGSe solar cells through direct *QE* and absolute EL measurements.

In both analyses, we found stronger broadening and higher radiative losses in absorbers that contain a gradient. CL analysis indicates for one sample, that the gradient might go along with stronger lateral fluctuations. We also quantified $\Delta\mu$ values with two different methods: linear fit to the high energy slope of Planck's generalized law ($Eq.\,20$) and adding up the overall $\Delta\mu$ losses ($Eq.\,6$ and $Eq.\,12$). Both methods result in nearly identical values for $\Delta\mu$, indicating that the loss analysis is complete. Finally, we investigated 19 different CIGSe samples from our laboratory. We observed that for our samples the dominant effect contributing to absorptance edge broadening is the presence of Ga gradient. Overall, our experimental results show that the presence of Ga gradient can lead to 5 to 16 meV higher radiative losses compared to non-graded samples. The meta-analysis and literature review also confirms that a gradient has a dominant effect on the broadening of *A(E)/Q(E)* onset.

## Experimental and methods:

*Sample preparation:* CISe and graded CIGSe absorbers were prepared using the 3-stage co-evaporation method [22, 58]. In this method, in the presence of Se atmosphere throughout the process, at the first stage, at the substrate temperature of 356°C, In and Ga were evaporated. In the second stage with the increase in the substrate temperature Cu was supplied to the absorbers. At the end of second stage, the absorbers have Cu rich composition (CGI ~ 1.2). Finally, at the last stage, the Cu flux was terminated, and again In and Ga were evaporated to adjust the overall composition to be Cu-poor. The maximum deposition temperature during the second and third stages was 580°C. For the CIGSe samples without band gap gradient, Cu-rich-off (CURO) co-evaporation method was employed [71]. Initially In, Ga, Cu and Se were co-evaporated simultaneously at the substrate temperature of 580 °C. Cu was supplied in excess to enhance the grain size and film quality. Then in the next stage, the supply of Cu was terminated, and only In, Ga, and Se were provided on the samples, resulting in an overall Cu-poor composition.

For transmittance/reflectance analysis, samples were prepared on clean soda lime glasses (SLG), while sample for solar cell fabrication were prepared on Mo-coated SLG substrates.

*Buffer layer deposition and solar cell fabrication:* All absorbers were covered by CdS. The CdS was deposited with chemical bath deposition method. Details regarding the CdS deposition can be found elsewhere [27]. Subsequently, for absorbers prepared on Mo, a ZnO and Al:ZnO double layer was sequentially sputtered as a window layer and the devices were finished by evaporating Al:Ni metallic grids on the surface.

*Photoluminescence (PL) measurements:* For all absorbers deposited on SLG, the PL measurements were conducted on CdS covered absorbers. The CdS prevents the surface degradation during the PL measurements [72]. A diode laser with a wavelength of 637 nm and beam radius of 1.3 mm was used as an excitation light source. The intensity of incident light was calibrated to 1 sun laser light flux. The PL signal was directed using two off-axis parabolic mirrors into fiber of 550 μm diameter, which directed the light into a spectrometer where an InGaAs detector was used to detect the signal. For the spectral calibration, a halogen lamp with the known light spectrum was employed. Absolute calibration is achieved measuring the laser with the spectrometer and with a power meter. Temperature of samples stays at ~ 296 K and it was independently measured using a thermal camera.

*Data smoothening:* To extract the broadening from the PL absorptance curve ($A^{PL}$), first we did smoothen of the absorptance ($A^{PL}$) spectra using Savitzky-Golay method. Then we took the derivative from smoothed curve to obtain broadening. (see **Fig S16** in **SI**)

*Electroluminescence (EL) measurements:* The EL measurements were conducted on solar cells using 4-probe contact configuration. A forward voltage bias was applied to the solar cell, and same as PL measurements, the EL emission was collected with two off-axis parabolic mirror and detected by an InGaAs detector. The EL spectrum was subsequently spectrally corrected using the halogen lamp. Moreover, we have performed absolute EL measurements to quantify the $Y_{EL}$. The injected current under one sun condition was applied to each solar cell which is equal to the short circuit current density of the solar cell. The normalized EL spectrum for F-BGG, G-NG, and H-NG solar cells can be found in **Fig S10** in the **SI**.

*Transmittance/reflectance measurements:* A photospectrometer is used to measure the transmittance and reflectance spectrums. This instrument is equipped with an integrating sphere and it covers photon wavelengths in the range of 300 -1800 nm. The light from the source is passed through a monochromator, then it reaches to the sample placed on the holder. After passing through the sample or reflecting from the surface, depending on the wavelength, the light is detected either by InGaAs or photomultiplier tubes detector.

*J-V analysis:* The J-V measurements were performed using a 4-probe contact configuration in a class AAA solar simulator. The solar simulator provides a simulated AM 1.5 spectrum. All the measurements were performed at 25°C (298 K) temperature. The solar cell area was independently measured using an optical microscope and the area is in the range of 0.21 cm$^2$ to 0.25 cm$^2$.

*External quantum efficiency (QE) measurements:* The *QE* measurements were performed on complete solar cells with a 4-probe contact configuration. The *QE* setup is equipped with a grating monochromator using chopped illumination of a halogen and a xenon lamp. Lock-in amplifier was used to measure the current of the solar cells and reference spectra based on calibrated Si and InGaAs detectors were used to calibrate the *QE* spectrum.

*Secondary ion mass spectroscopy (SIMS) measurements*: The SIMS measurements were carried out on a CAMECA SC-Ultra instrument. The films were analyzed using a $Cs^+$ low-impact energy beam (1keV) in the depth profiling mode. With such analytical conditions, the variation of elemental composition was acquired across the whole thickness of the films (2 – 2.5 µm). The intensity of the elements of interest (Cu, In, Ga and Se) was collected from a 60 µm area in diameter centered in a 250 µm × 250 µm scanned area.

*Cathodoluminescence measurements:* The CL measurements were conducted using an Attolight Allalin 4027 dedicated CL-SEM, equipped with an Oxford Instruments iDus InGaAs 1.7 detector. All CL maps were acquired at room temperature with an electron beam acceleration voltage of 5 kV, a probe current of 10 nA, and a 100 µm aperture. Both plan-view and cross-section maps were captured with a pixel exposure time of 250 ms. The open-source Python packages HyperSpy [73] and LumiSpy [74] were used for processing and analyzing the CL data.

*Meta-analysis and literature study:* In **Fig 9**, we present a meta-analysis of published data, showing the broadening parameter for various absorber technologies. The σ values were either directly extracted from reported values in previous studies or, in cases where they were not explicitly provided, obtained through digitization of the published *QE* curves, followed by broadening extraction.

# Acknowledgement

Financial support from the Luxembourgish Fonds National de la Recherche (FNR) in the framework of the projects TAILS, FULL, and REACH under the grant numbers C20/MS/14735144/TAILS, CORE/22/MS/17138895/FULL, and INTER/UKRI/20/15050982 is appreciated. Cambridge team would like to acknowledge funding from the Engineering and Physical Science Research Council (EPSRC) under EP/R025193/1. For the purpose of open access, the author has applied a Creative Commons Attribution 4.0 International (CC BY 4.0) license to any Author Accepted Manuscript version arising from this submission. ChatGPT, a language model developed by OpenAI in San Francisco, CA, USA, provided assistance in English language editing and improving clarity of our own written content. The whole text has been carefully modified and verified by the authors.

## Data availability

All data supporting this study will be available in Zenodo link.

## Supplementary information (SI)

The Supplementary Information (SI) is available.

## Authors Contribution

**Sevan Gharabeiki (Lead):** Writing Original Draft, Conceptualization, Visualization, Formal Analysis, Methodology, Investigation, Resources, Software. **Francesco Lodola:** Resources, Investigation, Writing Review And Editing, Software. **Tilly Schaaf**: Investigation, Visualization, Writing Review And Editing **Taowen Wang**: Resources, Writing Review And Editing. **Michele Melchiorre**: Investigation **Nathalie Valle**: Investigation, Resources, Writing Review And Editing. **Jeremy Niclout**: Investigation. **Manha Ali**: Investigation, Visualization. **Yucheng Hu**: Investigation, Visualization, Writing Review And Editing. **Gunnar Kusch**: Investigation, Visualization, Writing Review And Editing. **Rachel A.Oliver**: Resources, Supervision, Writing Review And Editing, Funding Acquisition **Susanne Siebentritt**: Conceptualization, Methodology, Resources, Supervision, Project Administration, Funding Acquisition, Writing Review And Editing.

# Appendix A. List of symbols and constants

| | *List of Symbols* | | | |
|---|---|---|---|---|
| *Symbol* | *Meaning (unit)* | *Symbol* | *Meaning (unit)* | |
| $E$ | Photon Energy (eV) | $J_0^{SQ}$ | Shockley-Queisser radiative saturation current density (A.cm$^{-2}$) | |
| $E_g$ | Band gap energy (eV) | $J_{SC}$ | Short circuit current density (A.cm$^{-2}$) | |
| $E_g^{loc}$ | Local Band gap energy (eV) | $J_{SC}^{SQ}$ | Shockley-Queisser short circuit current density (A.cm$^{-2}$) | |
| $\overline{E_g}$ | Mean band gap energy (eV) | $Y_{PL}$ | PL quantum yield | |
| $E_U$ | Urbach energy (eV) | $Y_{EL}$ | EL quantum yield | |
| $A(E)$ | Absorptance spectrum | $V_{OC}$ | Open circuit voltage (V) | |
| $A^{PL}(E)$ | Absorptance spectrum from PL | $\Delta\mu$ | Quasi Fermi level splitting (eV) | |
| $A^{dir}(E)$ | Absorptance spectrum from direct measurements | $V_{OC}^{SQ}$ | Shockley-Queisser $V_{OC}$ limit (V) | |
| $QE(E)$ | External Quantum efficiency spectrum | $qV_{OC}^{SQ}$ | Shockley-Queisser $\Delta\mu$ limit (eV) | |
| $QE^{EL}(E)$ | External Quantum efficiency spectrum from EL | $\Delta\mu^{rad}$ | Radiative $\Delta\mu$ limit (eV) | |
| $QE^{dir}(E)$ | External Quantum efficiency spectrum from direct measurements | $\delta\Delta\mu^{rad}$ | Radiative $\Delta\mu$ loss (eV) | |
| $\sigma_A$ | Absorptance edge broadening parameter (eV) | $\delta\Delta\mu^{Gen}$ | Generation $\Delta\mu$ loss (eV) | |
| $\sigma_g$ | Band gap fluctuation (inhomogeneities) parameter (eV) | $\delta\Delta\mu^{nr}$ | Non-radiative $\Delta\mu$ loss (eV) | |
| $\sigma_{QE}$ | External quantum efficiency broadening parameter (eV) | $V_{OC}^{rad}$ | Radiative $V_{OC}$ limit (V) | |
| $\Phi_{Sun}(E)$ | Sun spectrum AM 1.5 g [75] (1.cm$^{-2}$s$^{-1}$V$^{-1}$) | $\delta V_{OC}^{rad}$ | Radiative $V_{OC}$ loss (V) | |
| $\Phi_{BB}(E)$ | Black body emission spectrum (1.cm$^{-2}$s$^{-1}$V$^{-1}$) | $\delta V_{OC}^{SC}$ | Short circuit $V_{OC}$ loss (V) | |
| $F_0^{SQ}$ | Shockley-Queisser radiative saturation flux density (cm$^{-2}$s$^{-1}$) | $\delta V_{OC}^{nr}$ | Non-radiative $V_{OC}$ loss (V) | |
| $F_0^{rad}$ | Radiative saturation flux density (cm$^{-2}$s$^{-1}$) | $V_{in}$ | Internal voltage (V) | |
| $F_{Gen}^{Sun}$ | Sun generation Flux (cm$^{-2}$s$^{-1}$) | $\Phi_{PL}(E)$ | PL emission spectrum (1.cm$^{-2}$s$^{-1}$V$^{-1}$) | |
| $F_{Gen}^{SQ}$ | Shockley-Queisser Sun generation Flux (cm$^{-2}$s$^{-1}$) | $\Phi_{EL}(E)$ | EL emission spectrum (1.cm$^{-2}$s$^{-1}$V$^{-1}$) | |
| $\alpha$ | Absorption co-efficient (cm$^{-1}$) | $F_{Gen}^{Laser}$ | Laser generation flux (cm$^{-2}$s$^{-1}$) | |
| T | Temperature (K) | $F_{Laser}^{inc}$ | Laser incident flux (cm$^{-2}$s$^{-1}$) | |
| $J_0^{rad}$ | Radiative saturation current density (A.cm$^{-2}$) | $A(E_{laser})$ | Absorptance at laser photon energy | |

### List of constants

| | |
|---|---|
| q | Elementary charge |
| e | Euler's number i.e. exp(1) |
| h | Planck's constant |
| $k_B$ | Boltzmann constant |
| c | Light velocity in vacuum |

## Appendix B. Generalized detailed balance model for optical measurements

The radiative $\Delta\mu$ limit is provided by **Eq. 2** in **Table 1** (in the main text) was introduced previously [4].

It is the $\Delta\mu$ that is obtained, when there is no non-radiative recombination. In that case the integrated emission flux ($F^{rad}$) is equal to the sun generation flux:

$$F^{rad} = F_{gen}^{Sun} \qquad (Eq. B1)$$

Thus we have:

$$\int_0^\infty A(E)\Phi_{BB}(E)\exp\left(\frac{\Delta\mu^{rad}}{kT}\right)dE = \int_0^\infty A(E)\Phi_{Sun}dE \qquad (Eq. B2)$$

Which leads directly to **Eq. 2**.

Rau et al. [7] used a mathematical expansion to calculate the $V_{OC}$ loss components separately in the solar cells by using generalized detailed balance limit for optoelectrical analysis. Using similar approach here starting with the **Eq. 2** and expanding it we write:

$$\Delta\mu^{rad} = k_B T \ln\left(\frac{\int_{E_g}^\infty \Phi_{Sun}dE}{\int_{E_g}^\infty \Phi_{BB}dE} \times \frac{\int_{E_g}^\infty \Phi_{BB}dE}{\int_0^\infty A(E)\Phi_{BB}dE} \times \frac{\int_0^\infty A(E)\Phi_{Sun}dE}{\int_{E_g}^\infty \Phi_{Sun}dE}\right) \qquad (Eq. B3)$$

$$= k_B T \ln\left(\frac{F_{Gen}^{SQ}}{F_0^{SQ}} \times \frac{F_0^{SQ}}{F_0^{rad}} \times \frac{F_{Gen}^{Sun}}{F_{Gen}^{SQ}}\right) \quad (Eq.\,B4)$$

$$= qV_{OC}^{SQ} - \delta\Delta\mu^{rad} - \delta\Delta\mu^{Gen} \quad (Eq.\,B5)$$

**Eq. B5** is same **Eq. 6** in the main text (without non-radiative loss component). As explained previously, here $F_{Gen}^{SQ}$ and $F_0^{SQ}$ are the SQ generation flux and SQ saturation flux respectively. The first term in the logarithm argument (**Eq. B3** and **Eq. B4**) gives us the SQ-$\Delta\mu$ limit for a defined band gap $E_g$. The second and third term illustrate the relative $\Delta\mu$ losses with respect to the SQ limit of this $E_g$. The second term can be defined as radiative $\Delta\mu$ loss:

$$\delta\Delta\mu^{rad} = -k_B T \ln\left(\frac{F_0^{SQ}}{F_0^{rad}}\right) \quad (Eq.\,B6)$$

And the last term as generation $\Delta\mu$ loss:

$$\delta\Delta\mu^{Gen} = -k_B T \ln\left(\frac{F_{Gen}^{Sun}}{F_{Gen}^{SQ}}\right) \quad (Eq.\,B7)$$

The $\delta\Delta\mu^{Gen}$ is due to incomplete absorption (i.e., *A(E) < 1* for *E > $E_g$*), while the $\delta\Delta\mu^{rad}$ is due to the broadening of *A(E)* spectra.

# Appendix C. Simulations for error function model for *A(E)*

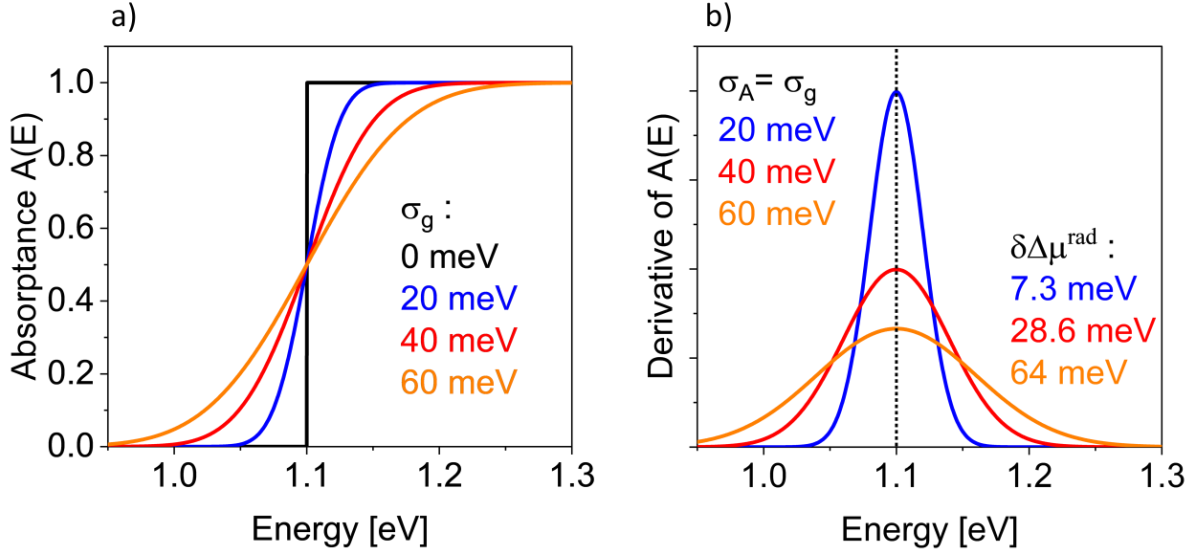

**Fig C1. a)** *A(E)* spectra modelled with error function model using ***Eq. 17***, with $\sigma_g$ of 0, 20, 40 and 60 meV. **b)** First derivative of simulated *A(E)* along with the radiative loss $\delta\Delta\mu^{rad}$ for each $\sigma_g$.

In **section 3** in the main text, we mentioned that the *A(E)* spectrum can be simply modelled as an error function like spectrum (***Eq. 17***), in **Fig C1.a** we show the simulated *A(E)* spectra for $\sigma_g$ in the range of 0 meV to 60 meV, and in **Fig C1.b** we illustrate the first derivative of these spectra that shows Gaussian distribution with broadening same as $\sigma_g$ (i.e., $\sigma_g = \sigma_A$). The $\sigma_g = 0$ meV resembles the SQ-*A(E)* spectrum (i.e, step function). Using ***Eq. 3***, we calculated the radiative losses for the simulated spectra and the values are summarized in **Fig C1. b** inset.

Furthermore, we also compared our calculated values with the previously published reports. Rau et al. proposed following formula to extract the $\Delta\mu^{rad}$ limit based on definition of *A(E)* as an error function (as ***Eq. 17*** in the main text) [8] :

$$\Delta\mu^{rad} = \overline{E_g} - \frac{\sigma_g^2}{2k_BT} - kT\ln\left(\frac{J_{00}}{J_{SC}}\right) \quad (Eq.C1)$$

Where,

$$J_{00} = \frac{2\pi q}{h^3c^2}\left[2(k_BT)^3 + 2(k_BT)^2\overline{E_g} + k_BT\overline{E_g}^2 - 2\sigma_g^2\overline{E_g} - \sigma_g^2 k_BT + \frac{\sigma_g^4}{k_BT}\right] \quad (Eq.C2)$$

And J<sub>SC</sub> as mentioned in the main text:

$$J_{SC} = q\int_0^\infty A(E)\Phi_{Sun}dE \quad (Eq.C3)$$

The details of the derivation of these equation can be found in ref [8]. In some reports simplified equations are being used to estimate the $\Delta\mu^{rad}$, which involves the following approximations [9, 35, 68]:

$$qV_{OC}^{SQ} \approx \overline{E_g} - k_BT\ln\left(\frac{J_{00}}{J_{SC}}\right) \quad (Eq.C4)$$

Then,

$$\Delta\mu^{rad} \approx qV_{OC}^{SQ} - \frac{\sigma_g^2}{2k_BT} \quad (Eq.C5)$$

In **Fig. C2** we compared our numerically calculated values for $\Delta\mu^{rad}$ using $Eq.2$, with proposed method by Rau et al [8] (i.e., see $Eq.C1$) and simplified approximation $Eq.C5$. It can be seen that our calculations are completely in agreement with $\Delta\mu^{rad}$ values extracted from $Eq.C1$.

Furthermore, regarding approximation in $Eq.C5$, we show that this approach induces errors in $\Delta\mu^{rad}$ determination particularly when σ$_g$ is large. In fact, when the σ$_g$ = 0 the $Eq.C1$ would be same as SQ limit:

However, as the σ$_g$ gets larger, the right-hand side of $Eq.C5$ gets higher values than $qV_{OC}^{SQ}$, therefore by simply approximating it with $qV_{OC}^{SQ}$ could result in the underestimation of the $\Delta\mu^{rad}$ and an overestimation of the radiative loss.

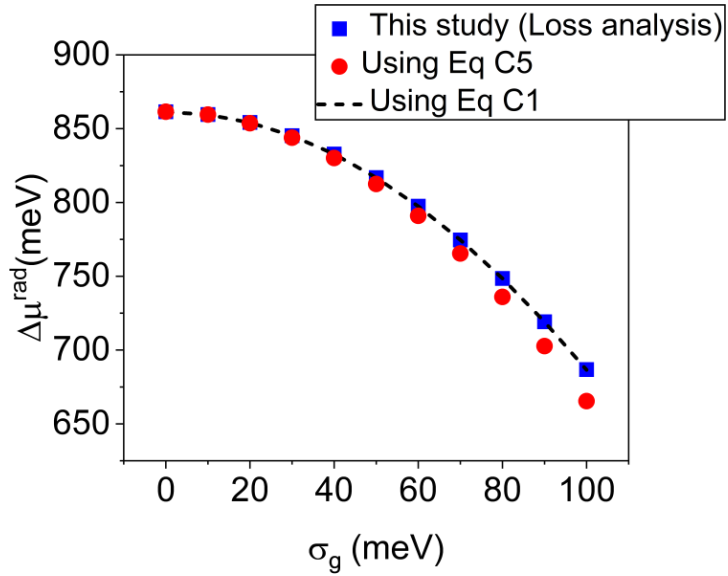

**Fig. C2** Radiative limit extracted from our numerical loss analysis $Eq.2$, from Rau analysis $Eq.C1$ [8] and approximation with $Eq.C5$.

## Appendix D. Effect of fluctuation and tail states on A(E)

**Fig D1** presents simulated *A(E)* spectra based on the absorption coefficient of a direct band gap material, exhibiting a square-root dependence above the band gap and Urbach decay behavior below it, combined with band gap fluctuations. In the absence of fluctuations ($\sigma_g$=0), the sub-bandgap *A(E)* would be zero, resulting in a discontinuous derivative. To address this special case, a small $E_U$=0.01 meV was introduced for simulations. The first derivatives and corresponding broadening for each curve is discussed in **SI note 4**.

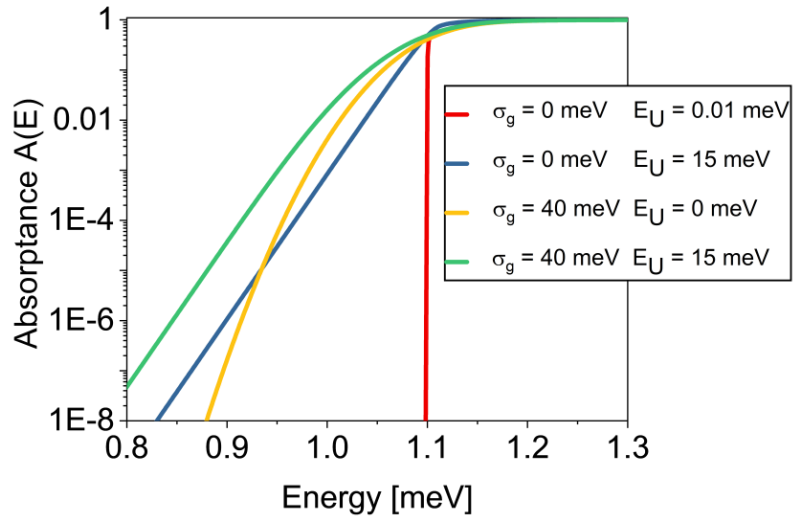

**Fig D1**. Simulated *A(E)* spectra by using absorption coefficient α of a real direct semiconductor, combined with Urbach tails and band gap fluctuations as explained in **section 3** in the main text. The separate effect of *A(E)* broadening and Urbach tails is clearly visible. The absorber is assumed 2 μm thick. For broadening analysis please see **SI note 4**.

# References:

[1] P. Wurfel, The chemical potential of radiation, Journal of Physics C: Solid State Physics, 15 (1982) 3967.
[2] P. Würfel, U. Würfel, Physics of solar cells: from basic principles to advanced concepts, John Wiley & Sons2016.
[3] S. Siebentritt, T.P. Weiss, M. Sood, M.H. Wolter, A. Lomuscio, O. Ramirez, How photoluminescence can predict the efficiency of solar cells, Journal of Physics: Materials, 4 (2021) 042010.
[4] S. Siebentritt, U. Rau, S. Gharabeiki, T.P. Weiss, A. Prot, T. Wang, D. Adeleye, M. Drahem, A. Singh, Photoluminescence assessment of materials for solar cell absorbers, Faraday Discuss, 239 (2022) 112-129.
[5] P. Caprioglio, M. Stolterfoht, C.M. Wolff, T. Unold, B. Rech, S. Albrecht, D. Neher, On the relation between the open‐circuit voltage and quasi‐fermi level splitting in efficient perovskite solar cells, Advanced Energy Materials, 9 (2019) 1901631.
[6] W. Shockley, H.J. Queisser, Detailed Balance Limit of Efficiency of p‐n Junction Solar Cells, Journal of Applied Physics, 32 (1961) 510-519.
[7] U. Rau, B. Blank, T.C.M. Müller, T. Kirchartz, Efficiency Potential of Photovoltaic Materials and Devices Unveiled by Detailed-Balance Analysis, Physical Review Applied, 7 (2017) 044016.
[8] U. Rau, J. Werner, Radiative efficiency limits of solar cells with lateral band-gap fluctuations, Applied physics letters, 84 (2004) 3735-3737.
[9] U.R. J. Mattheis, and J. Werner, Light absorption and emission on semiconductors with band gap fluctuations - a study on Cu(In,Ga)Se2 thin films, J. Appl. Phys, 101 113519.
[10] J.H. Werner, J. Mattheis, U. Rau, Efficiency limitations of polycrystalline thin film solar cells: case of Cu(In,Ga)Se2, Thin Solid Films, 480-481 (2005) 399-409.
[11] M.H. Wolter, R. Carron, E. Avancini, B. Bissig, T.P. Weiss, S. Nishiwaki, T. Feurer, S. Buecheler, P. Jackson, W. Witte, S. Siebentritt, How band tail recombination influences the open‐circuit voltage of solar cells, Progress in Photovoltaics: Research and Applications, DOI 10.1002/pip.3449(2021) 702–712.
[12] S. Thomas, T. Bertram, C. Kaufmann, T. Kodalle, J.A. Márquez Prieto, H. Hempel, L. Choubrac, W. Witte, D. Hariskos, R. Mainz, Effects of material properties of band‐gap‐graded Cu (In, Ga) Se2 thin films on the onset of the quantum efficiency spectra of corresponding solar cells, Progress in Photovoltaics: Research and Applications, 30 (2022) 1238-1246.
[13] L. Krückemeier, U. Rau, M. Stolterfoht, T. Kirchartz, How to report record open‐circuit voltages in lead‐halide perovskite solar cells, Advanced energy materials, 10 (2020) 1902573.
[14] W. Shockley, W. Read Jr, Statistics of the recombinations of holes and electrons, Physical review, 87 (1952) 835.
[15] R.N. Hall, Electron-hole recombination in germanium, Physical review, 87 (1952) 387.
[16] T. Tiedje, E. Yablonovitch, G.D. Cody, B.G. Brooks, Limiting efficiency of silicon solar cells, IEEE Transactions on electron devices, 31 (1984) 711-716.
[17] M. Govoni, I. Marri, S. Ossicini, Auger recombination in Si and GaAs semiconductors: Ab initio results, Physical Review B—Condensed Matter and Materials Physics, 84 (2011) 075215.
[18] D. Matsakis, A. Coster, B. Laster, R. Sime, A renaming proposal:"The Auger–Meitner effect", Physics Today, 72 (2019) 10-11.
[19] U. Rau, Reciprocity relation between photovoltaic quantum efficiency and electroluminescent emission of solar cells, Physical Review B, 76 (2007) 085303.
[20] S. Gharabeiki, M.U. Farooq, T. Wang, M. Sood, M. Melchiorre, C.A. Kaufmann, A. Redinger, S. Siebentritt, Grain boundaries are not the source of Urbach tails in Cu (In, Ga) Se2 absorbers, Journal of Physics: Energy, 6 (2024) 035008.

[21] O. Ramírez, J. Nishinaga, F. Dingwell, T. Wang, A. Prot, M.H. Wolter, V. Ranjan, S. Siebentritt, On the Origin of Tail States and Open Circuit Voltage Losses in Cu (In, Ga) Se2, Solar RRL, 7 (2023) 2300054.
[22] A.M. Gabor, J.R. Tuttle, M.H. Bode, A. Franz, A.L. Tennant, M.A. Contreras, R. Noufi, D.G. Jensen, A.M. Hermann, Band-gap engineering in Cu (In, Ga) Se2 thin films grown from (In, Ga) 2Se3 precursors, Solar energy materials and solar cells, 41 (1996) 247-260.
[23] P. Jackson, R. Wuerz, D. Hariskos, E. Lotter, W. Witte, M. Powalla, Effects of heavy alkali elements in Cu(In,Ga)Se2 solar cells with efficiencies up to 22.6%, physica status solidi (RRL) – Rapid Research Letters, 10 (2016) 583-586.
[24] A. Chirilă, S. Buecheler, F. Pianezzi, P. Bloesch, C. Gretener, A.R. Uhl, C. Fella, L. Kranz, J. Perrenoud, S. Seyrling, Highly efficient Cu (In, Ga) Se2 solar cells grown on flexible polymer films, Nature materials, 10 (2011) 857-861.
[25] W. Witte, D. Abou‐Ras, K. Albe, G.H. Bauer, F. Bertram, C. Boit, R. Brüggemann, J. Christen, J. Dietrich, A. Eicke, Gallium gradients in Cu (In, Ga) Se2 thin‐film solar cells, Progress in Photovoltaics: Research and Applications, 23 (2015) 717-733.
[26] T. Feurer, P. Reinhard, E. Avancini, B. Bissig, J. Löckinger, P. Fuchs, R. Carron, T.P. Weiss, J. Perrenoud, S. Stutterheim, Progress in thin film CIGS photovoltaics–Research and development, manufacturing, and applications, Progress in Photovoltaics: Research and Applications, 25 (2017) 645-667.
[27] T. Wang, F. Ehre, T.P. Weiss, B. Veith‐Wolf, V. Titova, N. Valle, M. Melchiorre, O. Ramírez, J. Schmidt, S. Siebentritt, Diode factor in solar cells with metastable defects and Back contact recombination, Advanced Energy Materials, 12 (2022) 2202076.
[28] M.A. Scarpulla, B. McCandless, A.B. Phillips, Y. Yan, M.J. Heben, C. Wolden, G. Xiong, W.K. Metzger, D. Mao, D. Krasikov, CdTe-based thin film photovoltaics: Recent advances, current challenges and future prospects, Solar Energy Materials and Solar Cells, 255 (2023) 112289.
[29] X. Zheng, D. Kuciauskas, J. Moseley, E. Colegrove, D.S. Albin, H. Moutinho, J.N. Duenow, T. Ablekim, S.P. Harvey, A. Ferguson, Recombination and bandgap engineering in CdSeTe/CdTe solar cells, Apl Materials, 7 (2019) 071112.
[30] Y. Zhao, S. Chen, M. Ishaq, M. Cathelinaud, C. Yan, H. Ma, P. Fan, X. Zhang, Z. Su, G. Liang, Controllable double gradient bandgap strategy enables high efficiency solution‐processed kesterite solar cells, Advanced Functional Materials, 34 (2024) 2311992.
[31] H. Phirke, S. Gharabeiki, A. Singh, A. Krishna, S. Siebentritt, A. Redinger, Quantifying recombination and charge carrier extraction in halide perovskites via hyperspectral time-resolved photoluminescence imaging, APL Energy, 2 (2024) 016111.
[32] U. Rau, V. Huhn, B.E. Pieters, Luminescence analysis of charge-carrier separation and internal series-resistance losses in Cu (In, Ga) Se 2 solar cells, Physical review applied, 14 (2020) 014046.
[33] R. Scheer, H.-W. Schock, Chalcogenide photovoltaics: physics, technologies, and thin film devices, John Wiley & Sons2011.
[34] T. Kirchartz, U. Rau, Electroluminescence analysis of high efficiency Cu (In, Ga) Se2 solar cells, Journal of applied physics, 102 (2007) 104510.
[35] D. Abou-Ras, Microscopic origins of radiative performance losses in thin-film solar cells at the example of (Ag, Cu)(In, Ga) Se2 devices, Journal of Vacuum Science & Technology A, 42 (2024) 022803.
[36] L. Gütay, G. Bauer, Spectrally resolved photoluminescence studies on Cu (In, Ga) Se2 solar cells with lateral submicron resolution, Thin Solid Films, 515 (2007) 6212-6216.
[37] L. Gütay, C. Lienau, G.H. Bauer, Subgrain size inhomogeneities in the luminescence spectra of thin film chalcopyrites, Applied Physics Letters, 97 (2010) 052110.
[38] J. Keller, K. Kiselman, O. Donzel-Gargand, N.M. Martin, M. Babucci, O. Lundberg, E. Wallin, L. Stolt, M. Edoff, High-concentration silver alloying and steep back-contact gallium grading enabling copper indium gallium selenide solar cell with 23.6% efficiency, Nature Energy, DOI 10.1038/s41560-024-01472-3(2024).

[39] M. Nakamura, K. Yamaguchi, Y. Kimoto, Y. Yasaki, T. Kato, H. Sugimoto, Cd-free Cu (In, Ga)(Se, S) 2 thin-film solar cell with record efficiency of 23.35%, IEEE Journal of Photovoltaics, 9 (2019) 1863-1867.
[40] M.I. Alonso, M. Garriga, C. Durante Rincón, E. Hernández, M. León, Optical functions of chalcopyrite CuGa x In 1-x Se 2 alloys, Applied Physics A, 74 (2002) 659-664.
[41] P. Paulson, S. Stephens, W. Shafarman, Analysis of Cu (InGa) Se 2 Alloy Film Optical Properties and the Effect of Cu Off-Stoichiometry, MRS Online Proceedings Library, 865 (2004) 141-146.
[42] G. Rey, G. Larramona, S. Bourdais, C. Choné, B. Delatouche, A. Jacob, G. Dennler, S. Siebentritt, On the origin of band-tails in kesterite, Solar Energy Materials and Solar Cells, 179 (2018) 142-151.
[43] K. Orgassa, Coherent optical analysis of the ZnO/CdS/Cu (In, Ga) Se2 thin film solar cell, Shaker2004.
[44] M. Grundmann, Physics of semiconductors, Springer2010.
[45] Alex Redinger, José A. Márquez , Thomas Unold, Quantitative Imaging of Non-radiative Losses in Thin Film Solar Cells, In Preparation.
[46] G. Rey, C. Spindler, F. Babbe, W. Rachad, S. Siebentritt, M. Nuys, R. Carius, S. Li, C. Platzer-Björkman, Absorption coefficient of a semiconductor thin film from photoluminescence, Physical Review Applied, 9 (2018) 064008.
[47] D. Abou-Ras, S. Siebentritt, Erratum:"Microscopic origins of radiative performance losses in thin-film solar cells at the example of (Ag, Cu)(In, Ga) Se2 devices"[J. Vac. Sci. Technol. A 42, 022803 (2024)], Journal of Vacuum Science & Technology A, 43 (2025).
[48] M.H. Wolter, WOLTER, M. (2019). Optical investigation of voltage losses in high-efficiency Cu(In,Ga)Se2 thin-film solar cells [Doctoral thesis, Unilu - University of Luxembourg]. ORBilu-University of Luxembourg. https://orbilu.uni.lu/handle/10993/39611, DOI (2019).
[49] S. Gharabeiki, M. Melchiorre, S. Siebentritt, Influence of NaF and KF Post-Deposition Treatment on the Sub-Band Gap Absorption of Cu (In, Ga) Se 2 Absorber Layers, 2023 IEEE 50th Photovoltaic Specialists Conference (PVSC), IEEE, 2023, pp. 1-3.
[50] E. Daub, P. Wurfel, Ultralow values of the absorption coefficient of Si obtained from luminescence, Phys Rev Lett, 74 (1995) 1020-1023.
[51] T.P. Weiss, O. Ramírez, S. Paetel, W. Witte, J. Nishinaga, T. Feurer, S. Siebentritt, Metastable defects decrease the fill factor of solar cells, Physical Review Applied, 19 (2023) 024052.
[52] T.P. Weiss, F. Ehre, V. Serrano-Escalante, T. Wang, S. Siebentritt, Understanding performance limitations of Cu (In, Ga) Se2 solar cells due to metastable defects—a route toward higher efficiencies, Solar RRL, 5 (2021) 2100063.
[53] L. Gütay, G. Bauer, Local fluctuations of absorber properties of Cu (In, Ga) Se2 by sub-micron resolved PL towards "real life" conditions, Thin Solid Films, 517 (2009) 2222-2225.
[54] T. Kirchartz, U. Rau, Detailed balance and reciprocity in solar cells, physica status solidi (a), 205 (2008) 2737-2751.
[55] T. Kirchartz, A. Helbig, W. Reetz, M. Reuter, J.H. Werner, U. Rau, Reciprocity between electroluminescence and quantum efficiency used for the characterization of silicon solar cells, Progress in Photovoltaics: Research and Applications, 17 (2009) 394-402.
[56] M. Sood, A. Urbaniak, C. Kameni Boumenou, T.P. Weiss, H. Elanzeery, F. Babbe, F. Werner, M. Melchiorre, S. Siebentritt, Near surface defects: Cause of deficit between internal and external open‐circuit voltage in solar cells, Progress in Photovoltaics: Research and Applications, 30 (2022) 263-275.
[57] S. Hegedus, D. Desai, C. Thompson, Voltage dependent photocurrent collection in CdTe/CdS solar cells, Progress in Photovoltaics: Research and Applications, 15 (2007) 587-602.
[58] T. Wang, L. Song, S. Gharabeiki, M. Sood, A.J.M. Prot, R.G. Poeira, M. Melchiorre, N. Valle, A.-M. Philippe, S. Glinsek, Shifting the paradigm: a functional hole selective transport layer for chalcopyrite solar cells, Solar RRL, 8 2400212.

[59] A.J. Prot, M. Melchiorre, F. Dingwell, A. Zelenina, H. Elanzeery, A. Lomuscio, T. Dalibor, M. Guc, R. Fonoll-Rubio, V. Izquierdo-Roca, Composition variations in Cu (In, Ga)(S, Se) 2 solar cells: Not a gradient, but an interlaced network of two phases, APL Materials, 11 (2023) 101120.
[60] S. Siebentritt, L. Gütay, D. Regesch, Y. Aida, V. Deprédurand, Why do we make Cu(In,Ga)Se2 solar cells non-stoichiometric?, Solar Energy Materials and Solar Cells, 119 (2013) 18-25.
[61] S. Siebentritt, E. Avancini, M. Bär, J. Bombsch, E. Bourgeois, S. Buecheler, R. Carron, C. Castro, S. Duguay, R. Félix, E. Handick, D. Hariskos, V. Havu, P. Jackson, H.P. Komsa, T. Kunze, M. Malitckaya, R. Menozzi, M. Nesladek, N. Nicoara, M. Puska, M. Raghuwanshi, P. Pareige, S. Sadewasser, G. Sozzi, A.N. Tiwari, S. Ueda, A. Vilalta‐Clemente, T.P. Weiss, F. Werner, R.G. Wilks, W. Witte, M.H. Wolter, Heavy Alkali Treatment of Cu(In,Ga)Se2 Solar Cells: Surface versus Bulk Effects, Advanced Energy Materials, 10 (2020).
[62] M.A. Green, K. Emery, Y. Hishikawa, W. Warta, E.D. Dunlop, Solar cell efficiency tables (version 40), Progress in Photovoltaics: Research and Applications, 20 (2012) 606-614.
[63] M.A. Green, E.D. Dunlop, M. Yoshita, N. Kopidakis, K. Bothe, G. Siefer, X. Hao, J.Y. Jiang, Solar cell efficiency tables (version 65), Progress in Photovoltaics: Research and Applications, 33 (2025) 3-15.
[64] M.A. Green, E.D. Dunlop, M. Yoshita, N. Kopidakis, K. Bothe, G. Siefer, D. Hinken, M. Rauer, J. Hohl‐Ebinger, X. Hao, Solar cell efficiency tables (Version 64), Progress in Photovoltaics: Research and Applications, 32 (2024) 425-441.
[65] K. Yin, J. Wang, L. Lou, X. Xu, B. Zhang, M. Jiao, J. Shi, D. Li, H. Wu, Y. Luo, Gradient bandgap enables> 13% efficiency sulfide Kesterite solar cells with open-circuit voltage over 800 mV, arXiv preprint arXiv:2404.00291, DOI (2024).
[66] M. Krause, S. Moser, C. Mitmit, S. Nishiwaki, A.N. Tiwari, R. Carron, Precise Alkali Supply during and after Growth for High‐Performance Low Bandgap (Ag, Cu) InSe2 Solar Cells, Solar RRL, 8 (2024) 2400077.
[67] J. Zhang, Z. Ma, Y. Zhang, X. Liu, R. Li, Q. Lin, G. Fang, X. Zheng, W. Li, C. Yang, Highly efficient narrow bandgap Cu (In, Ga) Se2 solar cells with enhanced open circuit voltage for tandem application, Nature Communications, 15 (2024) 10365.
[68] M. Krause, A. Nikolaeva, M. Maiberg, P. Jackson, D. Hariskos, W. Witte, J.A. Márquez, S. Levcenko, T. Unold, R. Scheer, Microscopic origins of performance losses in highly efficient Cu (In, Ga) Se2 thin-film solar cells, Nature communications, 11 (2020) 4189.
[69] S. Thomas, W. Witte, D. Hariskos, R. Gutzler, S. Paetel, C.Y. Song, H. Kempa, M. Maiberg, D. Abou‐Ras, Role of Ag Addition on the Microscopic Material Properties of (Ag, Cu)(In, Ga) Se2 Absorbers and Their Effects on Losses in the Open‐Circuit Voltage of Corresponding Devices, Progress in Photovoltaics: Research and Applications, 32 (2024) 930-940.
[70] T. Gokmen, O. Gunawan, T.K. Todorov, D.B. Mitzi, Band tailing and efficiency limitation in kesterite solar cells, Applied Physics Letters, 103 (2013) 103506.
[71] J. Kessler, C. Chityuttakan, J. Lu, J. Schöldström, L. Stolt, Cu (In, Ga) Se2 thin films grown with a Cu‐poor/rich/poor sequence: growth model and structural considerations, Progress in Photovoltaics: Research and Applications, 11 (2003) 319-331.
[72] F. Babbe, L. Choubrac, S. Siebentritt, Quasi Fermi level splitting of Cu-rich and Cu-poor Cu (In, Ga) Se2 absorber layers, Applied Physics Letters, 109 (2016).
[73] E.P. Francisco de la Peña, Vidar Tonaas Fauske, Jonas Lähnemann, Pierre Burdet, Petras Jokubauskas, Tom Furnival, Carter Francis, Magnus Nord, Tomas Ostasevicius, Katherine E. MacArthur, Duncan N. Johnstone, Mike Sarahan, Joshua Taillon, Thomas Aarholt, pquinn-dls, Vadim Migunov, Alberto Eljarrat, Jan Caron, … hyperspy/hyperspy: v2.1.1 (v2.1.1). Zenodo, DOI https://doi.org/10.5281/zenodo.12724131(2024).
[74] J. Lähnemann, J. Ferrer Orri, E. Prestat, H. Wiik Ånes, D.N. Johnstone, L. Migrator, N. Tappy, LumiSpy/lumispy: v0. 2.2, Zenodo, DOI 10.5281/zenodo.7747350(2023).
[75] ASTM G173-03 Reference Spectra http://rredc.nrel.gov/solar/spectra/am1.5/, NREL.

# The effect of a band gap gradient on the radiative losses in the open circuit voltage of solar cells
# Supporting information

**Sevan Gharabeiki** [1], Francesco Lodola [1], Tilly Schaaf [1,2], Taowen Wang [1], Michele Melchiorre [1], Nathalie Valle [3], Jérémy Niclout [3], Manha Ali [4], Yucheng Hu [4], Gunnar Kusch [4], Rachel A. Oliver [4] and Susanne Siebentritt [1]

1. Department of Physics and Material Science, University of Luxembourg, 41, rue du Brill, L-4422, Belvaux, Luxembourg

2. On leave from Trinity College Dublin, College Green, Dublin 2 Ireland

3. Luxembourg Institute of Science and Technology (LIST), 41, rue du Brill, L-4422, Belvaux, Luxembourg

4. Department of Materials Science and Metallurgy, University of Cambridge, 27 Charles Babbage Road, Cambridge CB3 0FS, United Kingdom

Email: sevan.gharabeiki@uni.lu, susanne.siebentritt@uni.lu

## SI note 1. Determination of $Y_{PL}$ through PL measurements

In our study, we perform absolute PL measurements to determine $Y_{PL}$, and the incident laser flux ($F_{Laser}^{inc}$) is calibrated to ~1 sun SQ generation flux (i.e., $F_{Laser}^{inc} \approx F_{Gen}^{SQ} = \int_{Eg}^{\infty} \Phi_{Sun} dE$). In this case for laser generation flux ($F_{Laser}^{Gen}$) we can write:

$$F_{Laser}^{Gen} = A(E_{laser}) F_{Laser}^{inc} \qquad (Eq.\,S1)$$

Here $A(E_{laser})$ is the value of $A(E)$ at the laser beam energy. In our study, we use a laser with an energy of 1.94 eV corresponding to a wavelength of 637 nm for our measurements. For our samples deposited on SLG, we measure the *A(E)* spectra using transmittance/reflectance measurements (See **Fig S1**). And we can determine the $A(E_{laser})$ with these measurements.

The monochromatic laser light, with a specific *A(E)* value at this energy, can result in a different absorbed photon flux compared to the solar spectrum, which spans a broad range of photon energies. However, here we show that this discrepancy is negligible.

Let's take B-NG as an example, under our experimental condition. The incident laser photon flux for this sample was calibrated to ~ $2.62 \times 10^{17}$ ($\frac{photons}{cm^{-2}s^{-1}}$). At the 637 nm the $A(E_{laser})$ value is 0.91. This would result in the generation flux ($F_{Gen}^{laser}$) of $2.38 \times 10^{17}$ ($\frac{photons}{cm^{-2}s^{-1}}$).

On the other hand, under sun spectrum for generation flux $F_{Gen}^{Sun}$ we can write:

$$F_{Gen}^{Sun} = \int_0^\infty A(E)\Phi_{Sun} dE \qquad (Eq.S2)$$

The $\Phi_{Sun}$ here is the AM1.5 solar spectrum according to [3]. Thus, by solving this integral (**Eq.S2**) for sample B-NG, we get to the values ~ $2.35 \times 10^{17}$ ($\frac{photons}{cm^{-2}s^{-1}}$).

Overall, we can write:

$$\frac{F_{Gen}^{Laser}}{F_{Gen}^{Sun}} \approx 1.01 \qquad (Eq.S3)$$

This small discrepancy would induce 0.3 meV error in the $\delta\Delta\mu^{nr}$ determination which is almost negligible.

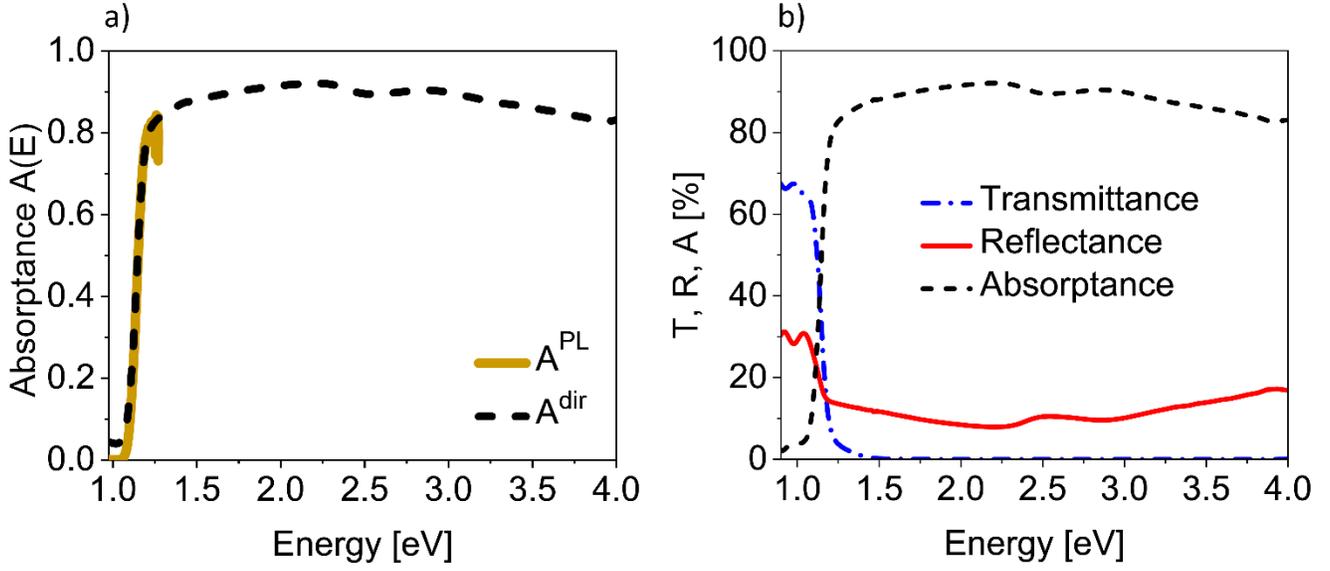

**Fig S1. a)** Full range *A(E)* spectra of B-NG sample. **b)** Transmittance, Reflectance and Absorptance (*A(E)*) measurements for B-NG sample.

## SI note 2. *A(E)* and *QE(E)* correlation

The *QE(E)* is related to *A(E)* spectra through the charge collection function [4]:

$$QE(E) = \int_0^d G(E,z)\eta(z)dz \qquad (Eq.\,S4)$$

Here $d$ is the thickness of absorber, and $G(E,z)$ is the generation function, which is depth dependent (i.e, depends on z) and also it is function of photons energy. The collection function $\eta(z)$ is only depth dependent. For $G(E,z)$ we can write [5]:

$$G(E,z) = (1 - R(E))\alpha(E)\exp(-\alpha z) \qquad (Eq.S5)$$

Where the $\alpha$ is the absorption coefficient of the material, and R is the surface reflectance.

If we assume that all carriers are collected through the junctions. The integral over the thickness of the material would be:

$$QE(E) = (1-R)(1-\exp(-\alpha d)) \quad if \ \eta(z) = 1 \qquad (Eq.S6)$$

Here the right-hand side of the *Eq. S6* is the A(E) spectra according to Beer-Lamberts law [6].

# SI note 3. Effect of $\sigma_g$, $E_U$ and thickness (*d*) on *A(E)*

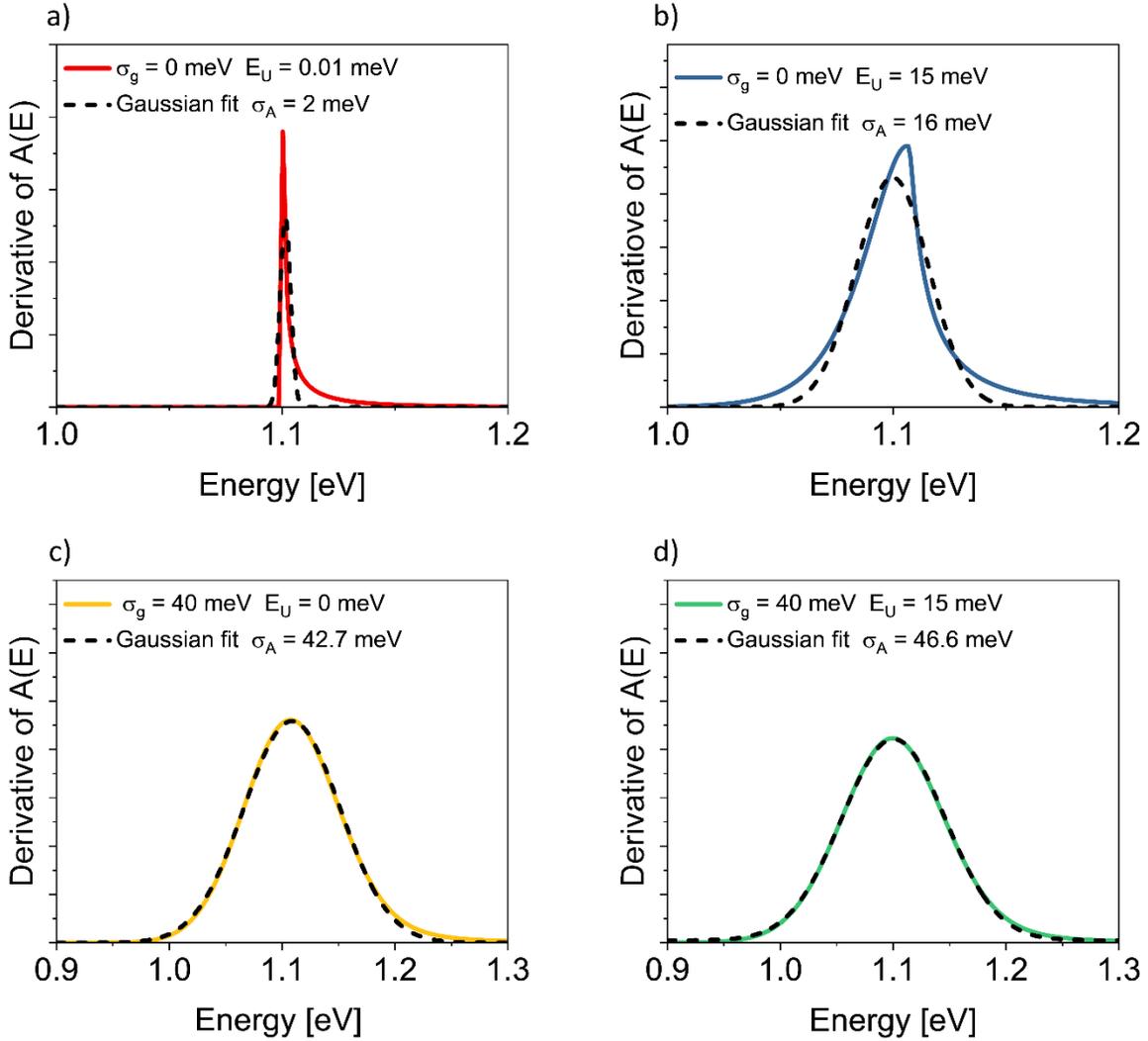

**Fig S2**. First derivative and corresponding Gaussian fit to simulated *A(E)* spectra with **a)** $\sigma_g = 0$ meV, $E_U = 0.01$ meV. **b)** $\sigma_g = 0$ meV, $E_U = 15$ meV. **c)** $\sigma_g = 40$ meV, $E_U = 0$ meV. **d)** $\sigma_g = 40$ meV, $E_U = 15$ meV. the $\sigma_A$ in each case is extracted from broadening parameter of Gaussian fit.

In Appendix D in the main text we presented the simulated the effect of fluctuation and Urbach tails on the final shape of A(E) spectra (see **Fig D1** in **Appendix D**). In **Fig S2** we show first derivative of *A(E)* for each case, along with the corresponding Gaussian fits. For each case the broadening of $A(E)$ onset and $\sigma_A$ is shown. When there are no fluctuations and $E_U$ is negligible (**Fig S2. a**), the square-root behavior alone still contributes to broadening of $A(E)$. Introducing $E_U$=15 meV, further increases this broadening (**Fig S2. b**), with an even greater effect observed when both fluctuations and combined fluctuations with tail states are present (**Fig S2. c and d**).

Furthermore, as we mentioned in the **section 3** in the main text, the broadening of the absorption edge can be also influenced by thickness of absorbers.

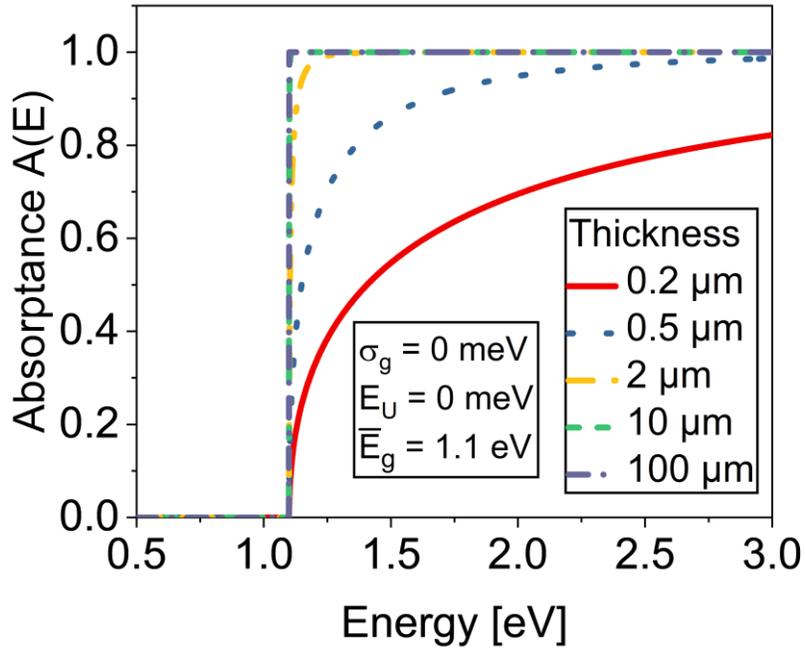

**Fig S3** The simulated $A(E)$ spectra for different thicknesses ($d$). with considering square root behavior of a above the band gap and $E_U = 0$ meV and $\sigma_g = 0$ meV.

In For simplicity we consider $\sigma_g = 0$ meV and $E_U = 0$ meV, and we use a square root alpha as defined in the **Eq. 14** in the main text. Here we do not show derivative, thus we used $E_U = 0$ meV for simulations.

The *A(E)* spectra for the absorbers with increasing thickness in the range of 0.2-100 μm are displayed in **Fig S3**. We see that with increasing thickness the *A(E)* spectrum gets closer to step function (i.e, SQ limit). And at small thicknesses the *A(E)* spectra shows increasing broadening.

## SI note 4. Secondary ion mass spectrometry (SIMS) measurement

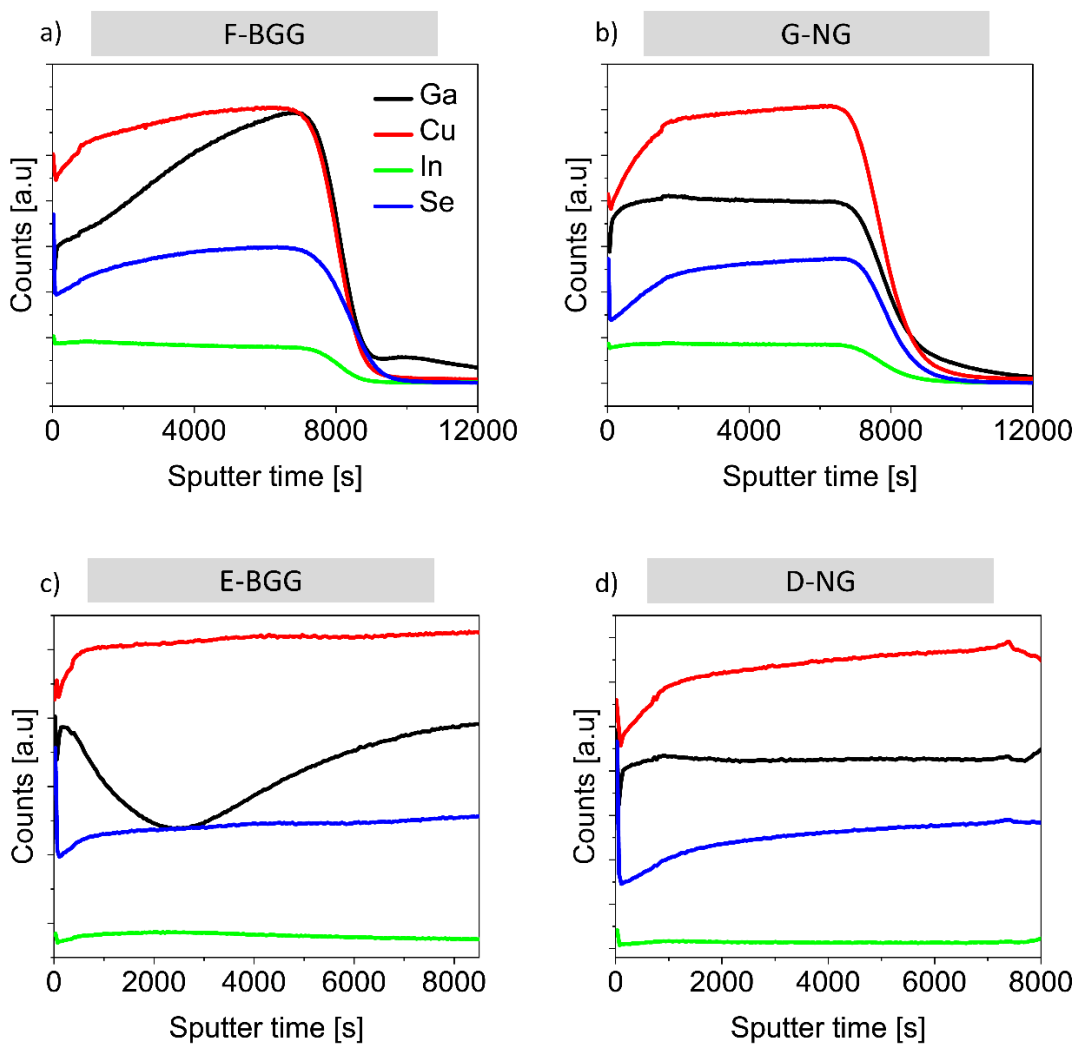

**Fig S4.** SIMS profile of elements Cu, In, Ga and Se for **a)** F-BGG, **b)** G-NG, **c)** D-NG and **d)** E-BGG samples. The surface of the film is located at sputter time = 0s.

# SI note 5. Absorptance spectra of D-NG and E-BGG samples

Absorptance spectra of D-NG and E-BGG absorbers. $A^{PL}$ denote absorptance extracted from PL measurements, and $A^{dir}$ is the absorptance measured from transmittance and reflectance measurements. (See **Fig S5**)

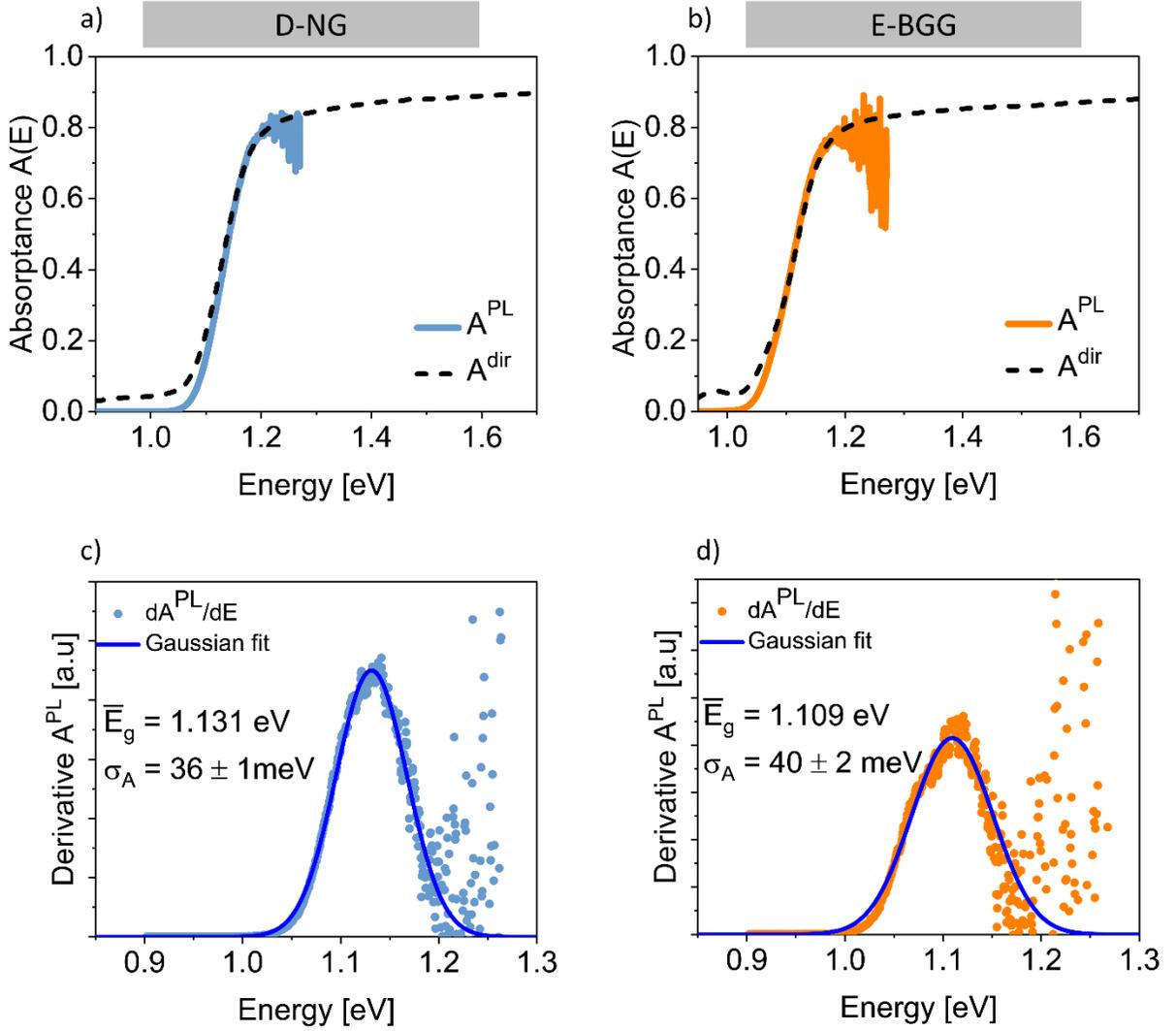

**Fig S5. a)**, **b)** Absorptance spectra of D-NG and E-BGG samples, **c)** and **d)** derivative of the $A^{PL}$ and gaussian fit for average band gap $\overline{E_g}$ and broadening $\sigma_A$ extaction. The derivatives are from smoothed curves.

## SI note 6. The first derivative of experimentally measured A(E)

In the **Fig S6**, we compare the first derivatives of *A(E)* spectra obtained from both PL measurements and direct measurements for B-NG sample. The inflection point, corresponding to the average band gap, is found to be at the same energy in both spectra. However, the curve derived from direct measurements exhibits lower values compared to those from PL analysis. This discrepancy arises primarily because the slope of $A^{dir}(E)$ at the inflection point is already affected by measurement limitations, leading to noise at low *A(E)* values. Additionally, on the low-energy side, interference peaks significantly influence the first derivative of $A^{dir}(E)$, causing their shape to deviate noticeably from the PL-derived spectra.

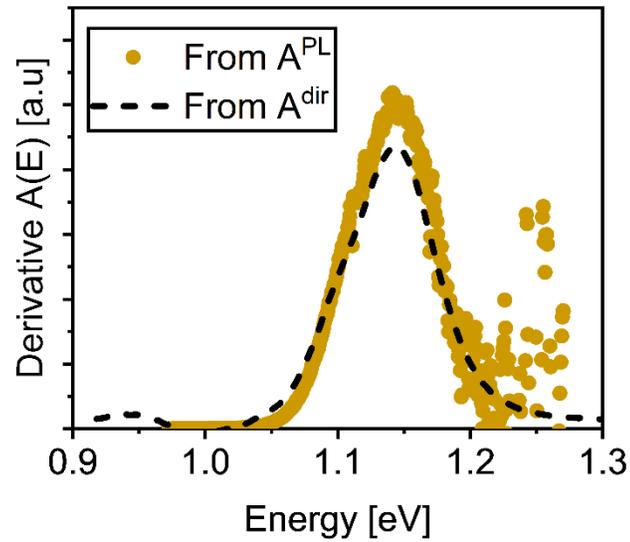

**Fig S6.** The fist derivate of *A(E)* spectra of B-NG sample extracted from PL measurements ($A^{PL}$) and direct measurements ($A^{dir}$).

# SI note 7. Incident intensity and $Y_{PL}$

**Fig S7. a** depicts the PL measurements of CIGSe B sample under increasing incident laser intensity. We can see in **Fig S7. b** that the $Y_{PL}$ increases with the increase in the incident laser intensity for all samples.

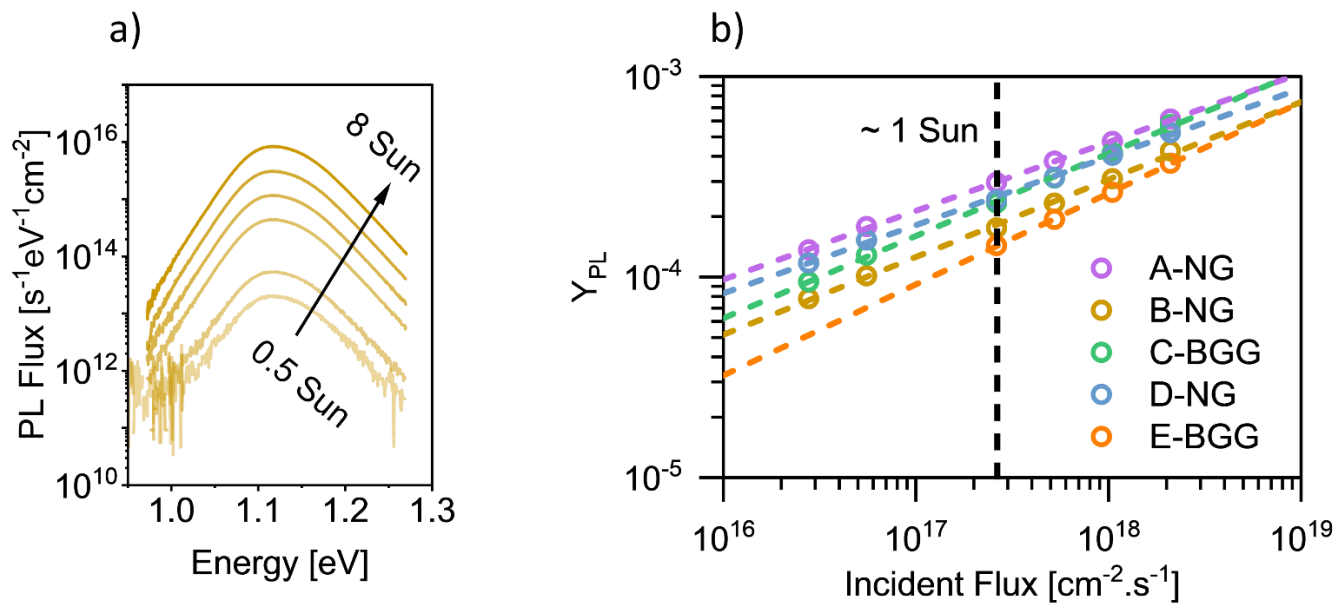

**Fig S7. a)** PL spectra of B-NG sample under increase in the laser intensity, **b)** $Y_{PL}$ of the samples deposited on glass with the increase in the intensity

## SI note 8. QFLS extracted from different methos.

The QFLS values reported in **Fig 5** are summarized in the following table.

**Supplementary Table S1**. QFLS values extracted from different methods reported in **Fig. 5** in the main text, values are in meV.

| Sample Name | QFLS extracted method | | |
|---|---|---|---|
| | Fit $A(E) = 1$ Eq. 20 | Fit $A(E) = A^{dir}(E)$ Eq. 20 | Eq.6 (loss analysis) |
| A-NG | 656.6 | 660.9 | 661.7 |
| B-NG | 655.3 | 661.9 | 662 |
| C-BGG | 671 | 678 | 677 |
| D-NG | 653 | 659.4 | 659.1 |
| E-BGG | 614 | 620.1 | 621 |

## SI note 9. Temperature from slope of the linear fit

Usually, the fitting range for the linear fit of the high energy slope of the PL spectrum is chosen at energies as high as possible, to "ensure" a constant $A(E)$ in the fitting range, but still it is important not to choose the range where the spectrum is affected by measurement noise. In **Fig S8** we show the absorptance spectra of CIGSe B-NG sample. We see that in the fitting range for QFLS extraction, the $A(E)$ spectrum does not yet level off and it continues to increase (**Fig S8. a**). In this case, the assumption $A(E) = 1$ results in the temperature extracted from slope of the fit to Planck's generalized law (**Eq.20** in main text) being higher than the actual measurement temperature. By using the measured $A^{dir}(E)$ spectrum instead of $A(E) = 1$, in the same fitting range, the fit temperature matches closely our measurement temperature (See **Fig S8. c**) However, it is important to mention that the extraction of temperature from fit can be challenging. Since at high energies and low PL intensity the slope can be influenced by measurement noise.

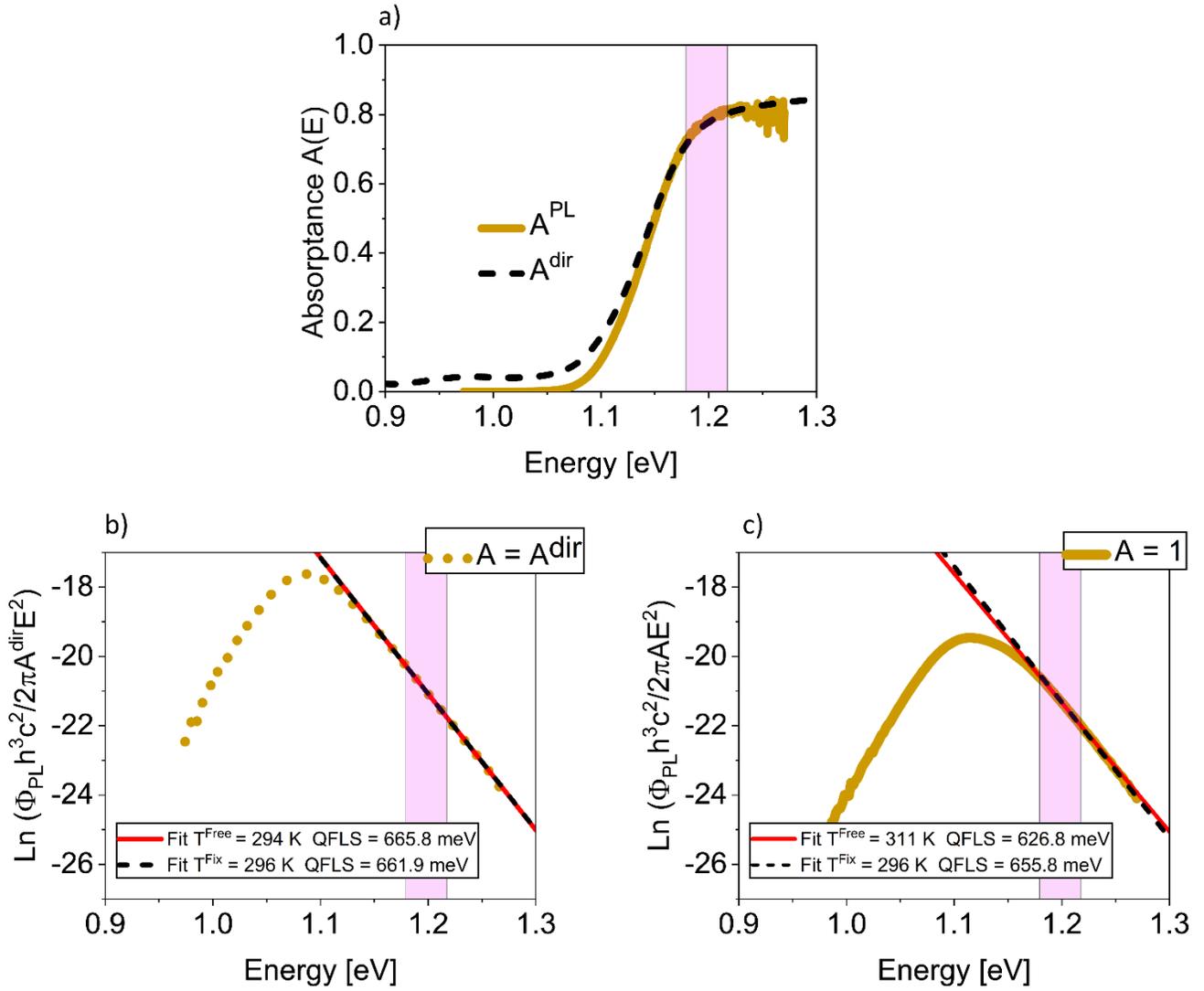

**Fig S8. a)** *A(E)* spectra of the B-NG sample. Pink area is the fitting range for the QFLS determination from the PL spectrum. **b)** Linear fits to the modified Planck's generalized law, assuming *A(E)* = 1, with both free-temperature and fixed-temperature approaches. **c)** Linear fits for QFLS extraction based on modified Planck's generalized law, considering $A(E)=A^{dir}$, using both free-temperature and fixed-temperature fits. Here, $A^{dir}$ indicated directly measured *A(E)* from photospectrometry. And the fitting range for QFLS is shown with highlighted region.

To clarify this issue, we have introduced the differential temperature ($T^{diff}$), which is simply the first derivative of modified Planck's generalized law (***Eq.20*** main text). With this approach at each point at the high energy side we can find the local slope and therefore the local extracted temperature. In **Fig S9. a**, we present the derivative of modified Planck's generalized law, assuming *A(E) = 1*. As shown, the slope is not constant within the fitting range, leading to an extracted temperature higher than the actual value. Conversely, when *A(E) = A$^{dir}$*, the slope becomes constant within the fitting range (**Fig S9. b**). However, it is important to note that the slope fluctuates, which may introduce a ± 2 K error in the extracted temperature from the linear fit, depending on the selected fitting range. A plot like in **Fig S9** is useful to check if the assumptions of the fit are met and the extracted QFLS can be trusted.

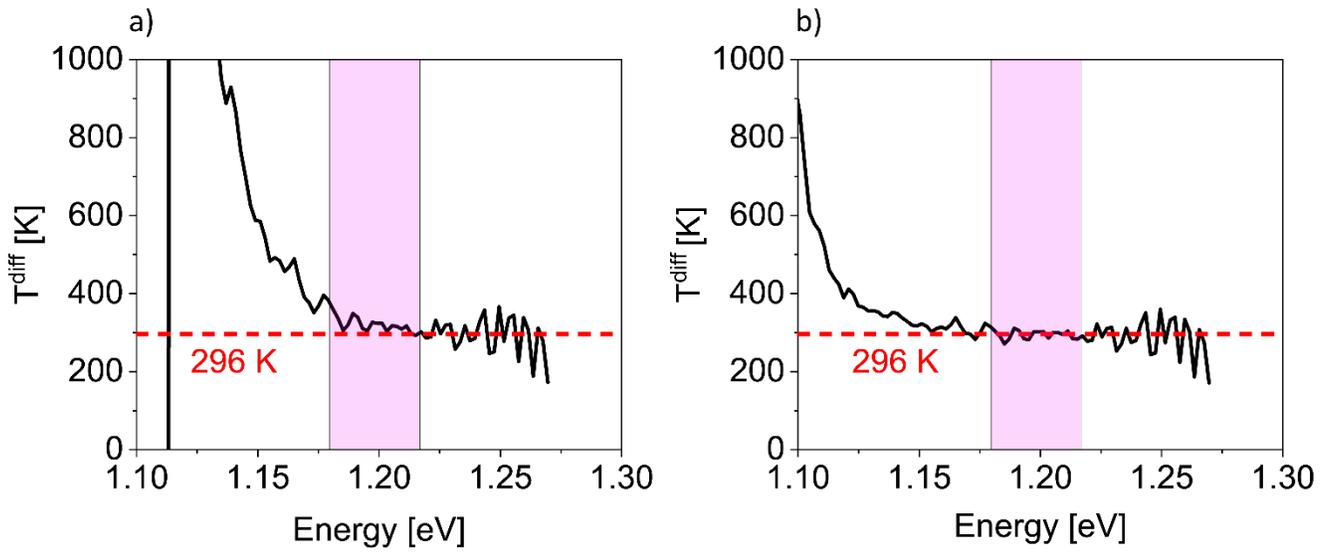

**Fig S9.** Differential temperature ($T^{diff}$) extracted for B-NG sample from first derivative of modified Planck's generalized law, by considering **a)** *A(E)* =1 and **b)** *A(E) = A$^{dir}$*. The fitting range for QFLS extraction is same as **Fig S8**.

# SI note 10. Normalized EL Spectra of F-BGG, G-NG and H-NG samples

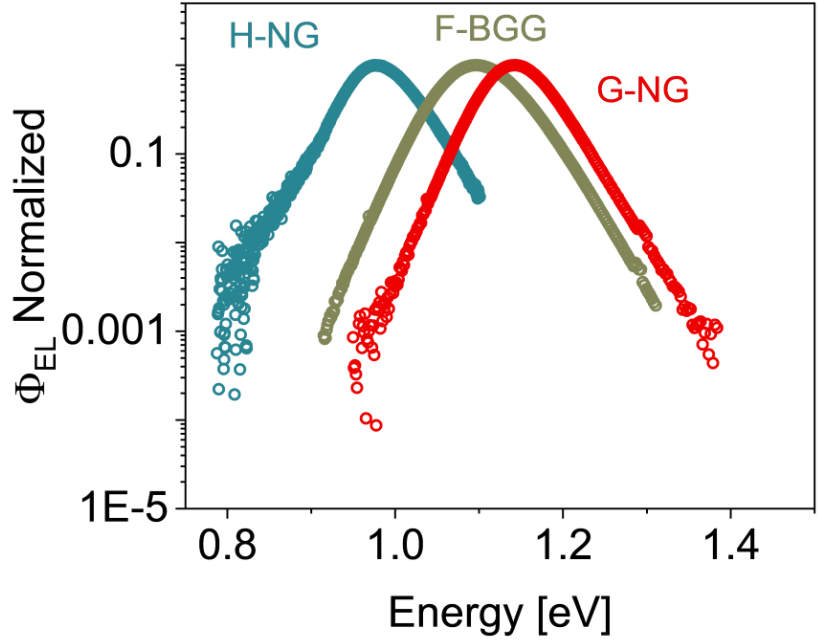

**Fig S10.** Electroluminesence spectra of F-BGG, G-NG and H-NG samples.

# SI note 11. *A(E)* and *QE* shape

As an example for sample F-BGG, in **Fig S11. a,** we show that $A^{PL}$ and $QE^{EL}$ have similar shape at the low energy decay.

Since both *A(E)* and *Q(E)* have similar decay shape, for Urbach Energy extraction we can use either $QE^{EL}$ or $A^{PL}$. According to Beer-Lamberts law we can write:

$$A(E) = QE(E) = 1-exp(-\alpha d)$$

Here, α is the absorption coefficient and d is the sample thickness (here 2μm). From the exponential decay of α at energies lower than the band gap energy we can extract the Urbach energy $E_U$ **Fig S11. b**. More details regarding the Urbach energy extraction can be found elsewhere [6, 11-13].

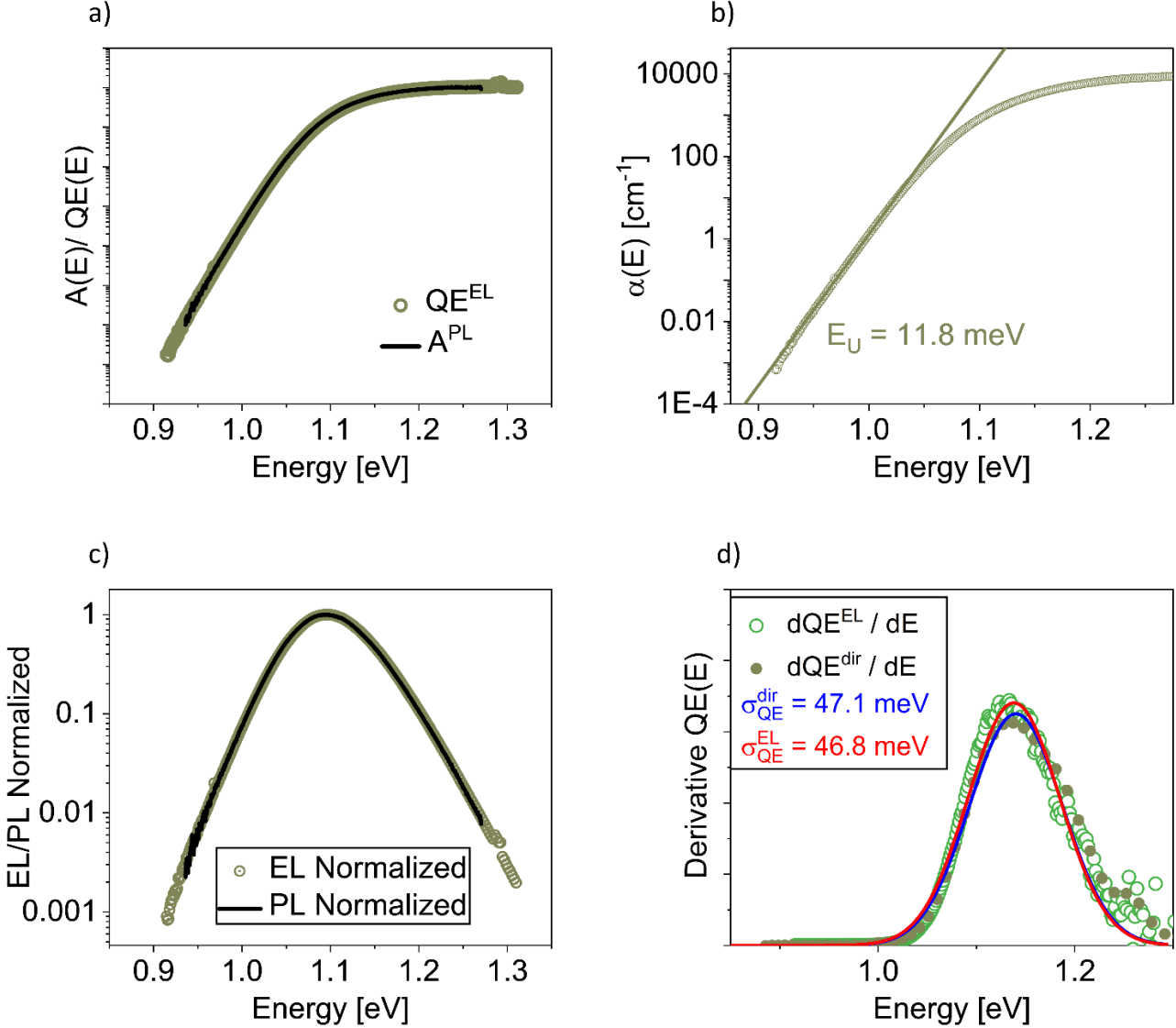

**Fig S11. a)** Absorptance extracted from PL ($A^{PL}$) and external quantum efficiency from EL measurements ($QE^{EL}$) for F-BGG sample. The $A^{PL}$ is not absolute and is rescaled to $QE^{EL}$. **b)** Urbach energy obtained from exponential decay of absorption coefficient for energy ranges below the band gap energy. **c)** Normalized PL and EL measurements, the PL measurements were performed on CIGSe/CdS/Window configuration **d)** First derivative of $QE^{EL}$ and $QE^{dir}$ spectra along with the Gaussian fit and broadening parameter. (Only one fitting range is illustrated here, in the main text the $\sigma_{QE}$ is average from different Gaussian fit)

In **Fig S11. c,** we show the normalized PL and EL spectra which have similar shape. In **Fig S11. d**, we present the first derivatives of $QE^{dir}$ and $QE^{EL}$. By performing a Gaussian fit, we extracted the broadening for both cases and observed that $\sigma_{QE}$ exhibits good agreement for both derivatives. Here, we display a single Gaussian fit for clarity. In the main text, the broadening was determined using different fitting ranges, with the reported value representing the average value along with errors.

It is important to mention that original QE derivative slightly deviates from the Gaussian fit. in our fitting process we ensure that low energy decay fit perfectly with Gaussian, since high energy part can be affected by measurement noise, in particular for EL measurements, and in QE measurements by non-linear collection effects and interferences in the window layer.

## SI note 12. Device parameters of solar cells

Statistical distribution of the device parameters for solar cells investigated in this study is summarized in **Table S2**, **Table S3** and **Table S4**. The solar cell with highest efficiency on each absorber in highlighted.

**Supplementary Table S2**. Device parameters of F-BGG sample.

| Sample No | $V_{OC}$(mV) | $J_{SC}$(mA/cm$^{-2}$) | FF | Eff (%) |
|---|---|---|---|---|
| F-1 | 637 | 35 | 76.7 | 17.1 |
| F-2 | 642 | 34.3 | 78.9 | 16.9 |
| F-3 | 642 | 34.8 | 75.1 | 16.8 |
| F-4 | 642 | 32.9 | 76.8 | 16 |
| F-5 | 643 | 35.3 | 77.2 | 17.5 |
| **F-6** | **648** | **35.5** | **76.7** | **17.6** |
| F-7 | 643 | 33.5 | 75.9 | 16.4 |
| F-8 | 646 | 31.1 | 76.3 | 15.3 |

**Supplementary Table S3**. Device parameters of G-NG sample.

| Sample No | $V_{OC}$ (mV) | $J_{SC}$ (mA/cm$^{-2}$) | FF | Eff (%) |
|---|---|---|---|---|
| G-1 | 607 | 31.2 | 76 | 14.4 |
| G-2 | 606 | 33 | 74.6 | 14.9 |
| G-3 | 607 | 33.4 | 75.3 | 15.3 |
| G-4 | 607 | 30.8 | 75.2 | 14.1 |
| G-5 | 608 | 33.7 | 76.1 | 15.6 |
| G-6 | 604 | 33.7 | 75.8 | 15.4 |
| G-7 | 602 | 32.7 | 75.6 | 14.9 |
| G-8 | 604 | 30.6 | 75.3 | 13.9 |

**Supplementary Table S4**. Device parameters of H-NG sample.

| Sample No | $V_{OC}$ (mV) | $J_{SC}$ (mA/cm$^{-2}$) | FF | Eff (%) |
|---|---|---|---|---|
| H-1 | 474 | 40.8 | 71.1 | 13.7 |
| H-2 | 473 | 39.9 | 71.7 | 13.5 |
| H-3 | 460 | 38.6 | 70.8 | 12.5 |
| H-4 | 468 | 39.8 | 68.3 | 12.7 |
| H-5 | 472 | 36.2 | 71.1 | 12.1 |

# SI note 13. SEM and CL analysis

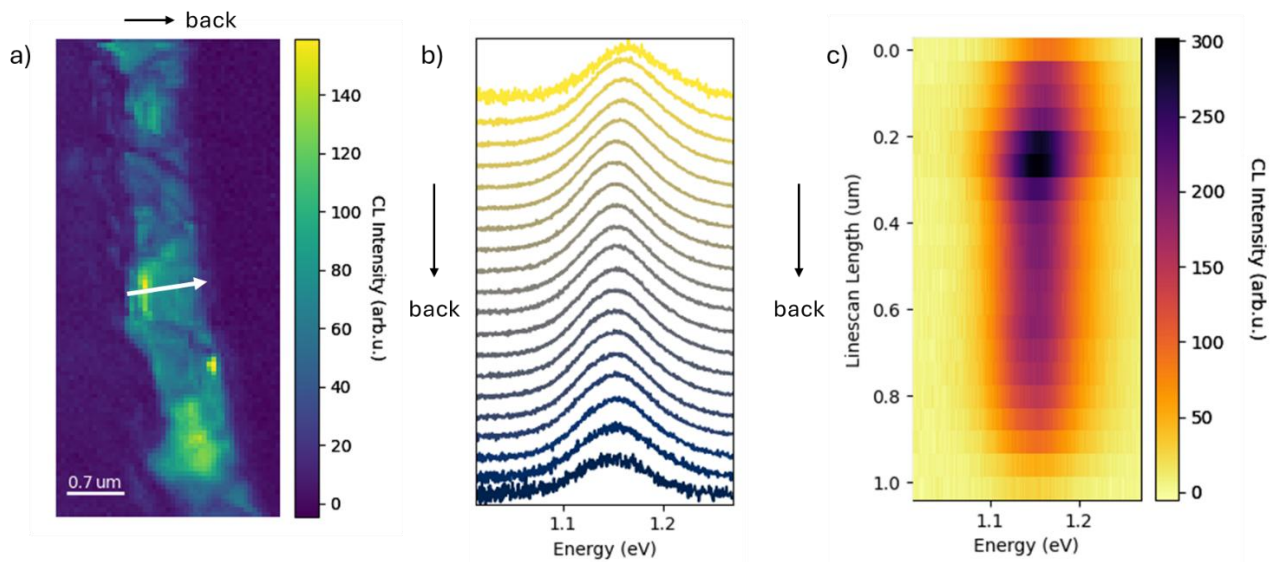

**Fig S12. a)** A panchromatic cross-section CL intensity map for the G-NG CIGSe sample, **b)** CL peak position along the line scan through the depth. heat map showing the CL emission maximum along the line scan.

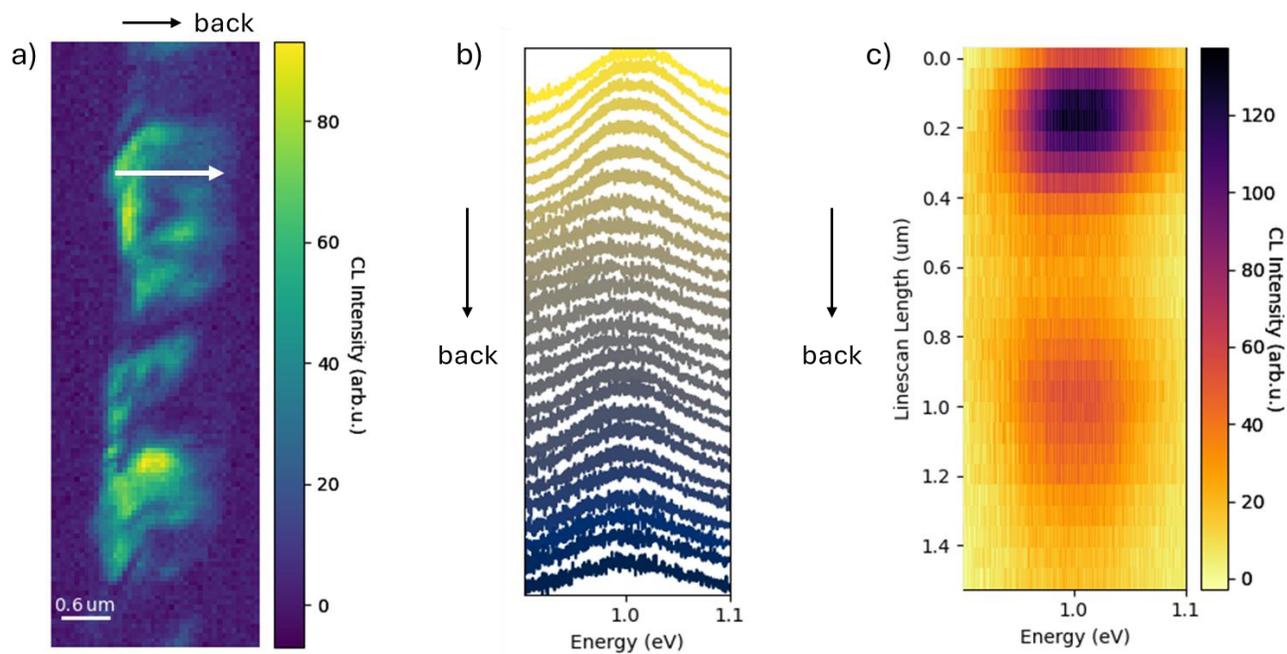

**Fig S13. a)** A panchromatic cross-section CL intensity map for the H-NG CIGSe sample, **b)** CL peak position along the line scan through the depth. heat map showing the CL emission maximum along the line scan.

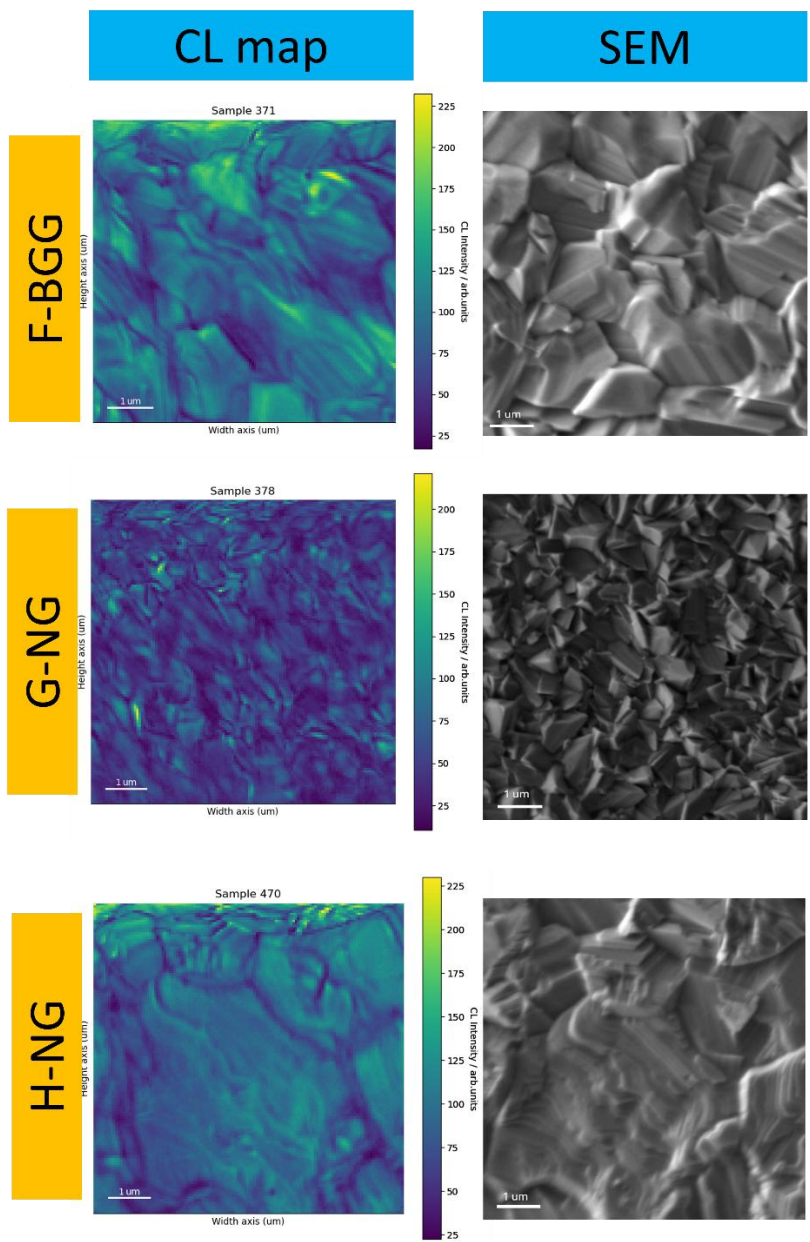

**Fig S14.** SE images showing the surface topography and panchromatic CL map for samples F-BGG, G-NG and H-NG, the SE images and CL maps have been acquired from the same area

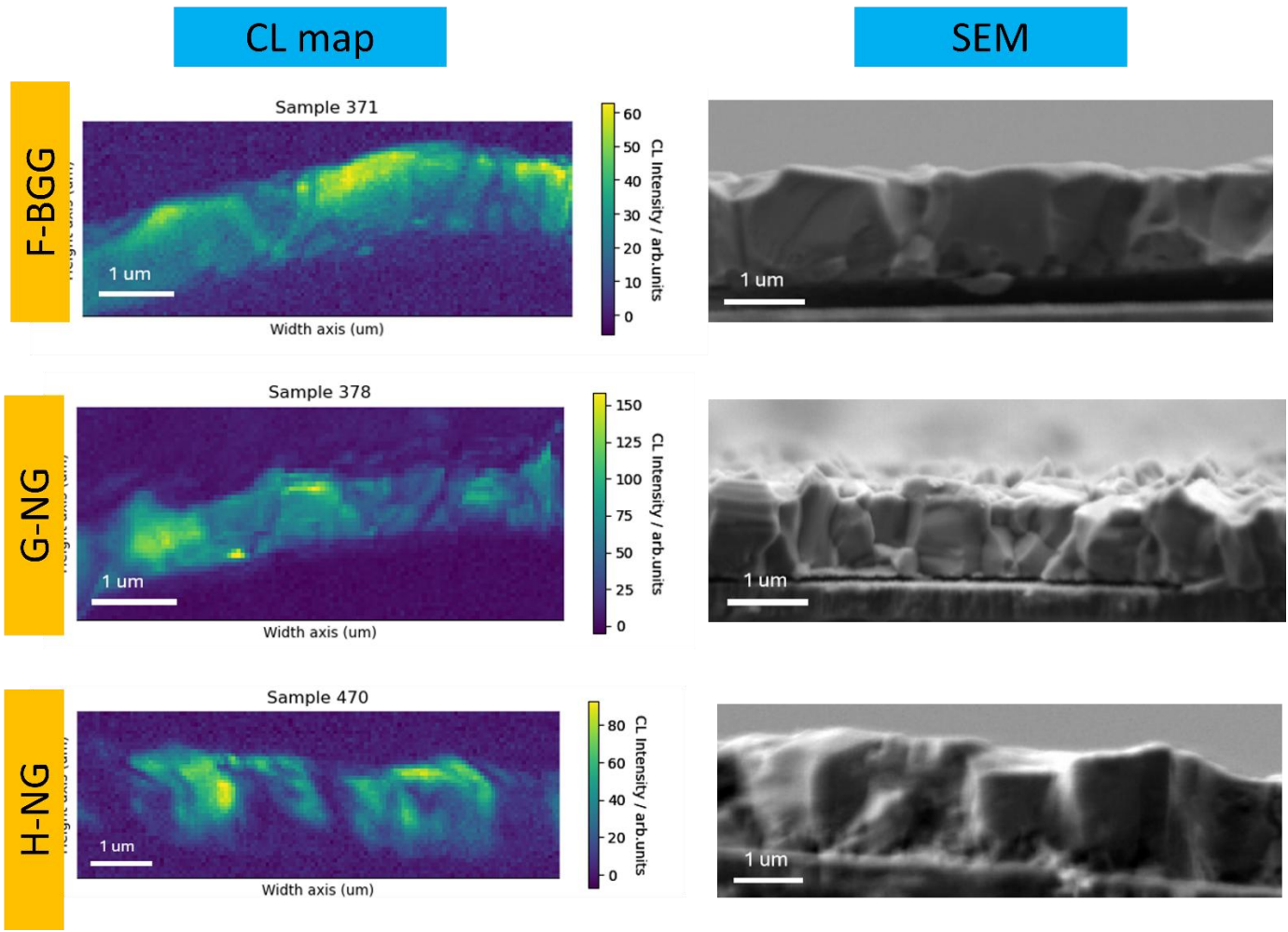

**Fig S15.** SE images of the cleaved cross-sections and panchromatic CL maps for samples F-BGG, G-NG and H-NG, the SE images and CL maps have been acquired from the same area, the difference in pattern between CL maps and SE images are charging-induced drift effect during long time measurement..

## SI note 14. Data smoothening Savitzky-Golay mathod

In **Fig S16** we present $A^{PL}$ curve smoothed by Savitzky-Golay method for B-NG sample.

Similar approach was also performed for other samples, and broadenings were extracted using first derivative of smoothed curves.

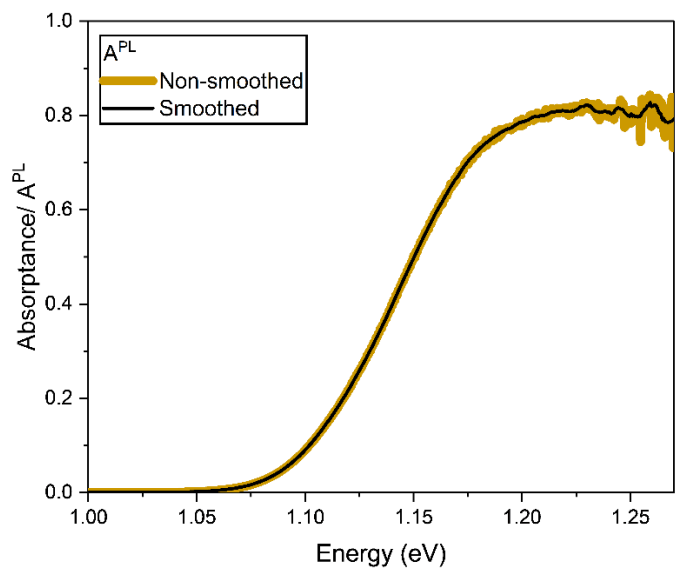

**Fig S16.** Smoothening of $A^{PL}$ curve for B-NG sample using Savitzky-Golay method.

## Supporting information references:

[1]  S. Siebentritt *et al.*, "Photoluminescence assessment of materials for solar cell absorbers," *Faraday Discuss,* vol. 239, no. 0, pp. 112-129, Oct 28 2022, doi: 10.1039/d2fd00057a.

[2]     U. Rau, B. Blank, T. C. M. Müller, and T. Kirchartz, "Efficiency Potential of Photovoltaic Materials and Devices Unveiled by Detailed-Balance Analysis," *Physical Review Applied,* vol. 7, no. 4, p. 044016, 2017, doi: 10.1103/PhysRevApplied.7.044016.
[3]     "https://www.nrel.gov/grid/solar-resource/spectra-am1.5.html." (accessed.
[4]     R. Scheer and H.-W. Schock, *Chalcogenide photovoltaics: physics, technologies, and thin film devices*. John Wiley & Sons, 2011.
[5]     P. Würfel and U. Würfel, *Physics of solar cells: from basic principles to advanced concepts*. John Wiley & Sons, 2016.
[6]     E. Daub and P. Wurfel, "Ultralow values of the absorption coefficient of Si obtained from luminescence," *Phys Rev Lett,* vol. 74, no. 6, pp. 1020-1023, Feb 6 1995, doi: 10.1103/PhysRevLett.74.1020.
[7]     U. Rau and J. Werner, "Radiative efficiency limits of solar cells with lateral band-gap fluctuations," *Applied physics letters,* vol. 84, no. 19, pp. 3735-3737, 2004.
[8]     U. R. J. Mattheis, and J. Werner, "Light absorption and emission on semiconductors with band gap fluctuations - a study on Cu(In,Ga)Se2 thin films," *J. Appl. Phys,* vol. 101, no. 11, p. 113519, 2007, doi: doi.org/10.1063/1.2721768.
[9]     M. Krause *et al.*, "Microscopic origins of performance losses in highly efficient Cu (In, Ga) Se2 thin-film solar cells," *Nature communications,* vol. 11, no. 1, p. 4189, 2020.
[10]   D. Abou-Ras, "Microscopic origins of radiative performance losses in thin-film solar cells at the example of (Ag, Cu)(In, Ga) Se2 devices," *Journal of Vacuum Science & Technology A,* vol. 42, no. 2, 2024.
[11]   S. Gharabeiki *et al.*, "Grain boundaries are not the source of Urbach tails in Cu(In,Ga)Se2 absorbers," *Submitted*.
[12]   M. H. Wolter, "Optical investigation of voltage losses in high-efficiency Cu (In, Ga) Se2 thin-film solar cells," PhD Thesis 2019.
[13]   L. Krückemeier, U. Rau, M. Stolterfoht, and T. Kirchartz, "How to report record open-circuit voltages in lead-halide perovskite solar cells," *Advanced energy materials,* vol. 10, no. 1, p. 1902573, 2020.